\documentclass[9pt,letterpaper,twoside,reqno,nosumlimits]{amsart}
\usepackage{amsmath,amssymb,amsthm}
\usepackage{graphicx}
\usepackage[colorlinks=True, citecolor=red, linkcolor=blue]{hyperref}
\usepackage{geometry}
\usepackage{xcolor}
\usepackage{subfigure}
\usepackage{algorithm}
\usepackage{algpseudocode}
\usepackage[foot]{amsaddr}

\usepackage{comment,color}
\usepackage[dvipsnames]{xcolor}
\patchcmd{\section}{\normalfont}{\normalfont\color{MidnightBlue}}{}{}
\patchcmd{\subsection}{\normalfont}{\normalfont\color{MidnightBlue}}{}{}

\begin{document}

\title{
Sparse stochastic digital twins can outperform exact ones
}
\author{
Zongren Zou\textsuperscript{a, 1}
\and Théo Bourdais\textsuperscript{a}
\and Ricardo Baptista\textsuperscript{b, 2}
\and Houman Owhadi\textsuperscript{a, 1, 2}
}

\footnotetext[1]{Corresponding authors: zzou@caltech.edu, owhadi@caltech.edu}
\footnotetext[2]{Joint senior authors}
\address{\textsuperscript{a} Department of Computing and Mathematical Sciences, California Institute of Technology, Pasadena, CA, USA}
\address{\textsuperscript{b} Department of Statistical Sciences, University of Toronto, Toronto, Canada}
\date{}

\begin{abstract}
Digital twins are often built under the assumption that greater model fidelity leads to better downstream performance. We show that this is not always the case: retaining weakly dependent or auxiliary degrees of freedom can increase the dimensionality of learning and control problems enough that a sparse stochastic surrogate outperforms even an exact high-dimensional model. To identify such structure, we introduce a framework for dependence discovery directly at the level of the full conditional law. The method combines conditional generative modeling with Gaussian-process distillation and kernel mode decomposition to quantify how strongly individual variables enter the conditional distribution, including through variability, tails, multimodality, and other features invisible to the conditional mean. The resulting dependence structure can be used both to construct sparse stochastic digital twins from data and to reduce known high-dimensional models for downstream tasks. In a modified Lunar Lander problem, an eight-dimensional sparse surrogate yields better control performance than the exact 109-dimensional simulator despite lower transition fidelity. A stochastic heat-equation example shows that the framework recovers sparse structure in a 102-dimensional data-driven setting and produces a more effective surrogate than one conditioned on all candidate inputs, while economic data demonstrate its use for identifying time-varying memory structure. These results show that model fidelity and downstream utility are distinct objectives, and that dependence discovery provides a practical route to sparse stochastic digital twins.
\end{abstract}

\maketitle

\section{Introduction}

Understanding how variables depend on one another is central to building accurate, efficient, interpretable, and useful digital twins of complex systems. Across physical, engineering, biological, and economic applications, this dependence often has substantial structure: only a subset of the available variables may be needed to describe the conditional behavior of a particular quantity of interest. Identifying this structure can reduce model complexity, improve interpretability by revealing the underlying dependence structure, and provide a principled basis for constructing sparse stochastic digital twins. This raises a broader question for digital twin modeling: does increasing model fidelity always improve downstream performance? We show that it does not: a sparse stochastic digital twin can outperform even an exact high-dimensional model. 

A wide range of methods address related structure-discovery problems, including variable selection in regression (e.g., \cite{mehmood2012review,lindsey2010variable,desboulets2018review,andersen2010variable,gu2019}), global and local sensitivity analysis (e.g., \cite{owen2013variance,kucherenko2016derivative,st1989analysis,mara2012variance,wahba2003introduction}), and sparse identification of governing equations \cite{brunton2016discovering}. These approaches differ substantially in their assumptions and objectives, but many widely used formulations characterize structure through deterministic or response-based representations of the target quantity, such as regression functions, sensitivity measures, governing equations, or conditional expectations. 

For stochastic digital twins, however, dependence may appear in the full conditional distribution without being visible in the conditional mean. This occurs naturally when unresolved variables, partial observations, multiscale dynamics, or intrinsic randomness make distributional uncertainty an essential part of the model. A candidate variable may alter variability, tail behavior, or multimodality while leaving the mean nearly unchanged. 
Such effects arise, for example, near critical transitions where fluctuations increase without large mean shifts \cite{scheffer2009early}, in coarse-grained and multiscale systems where unresolved variables induce state-dependent stochasticity \cite{givon2004extracting,chorin2002optimal,gao2024learning}, and in systems where rare events or downside risk are of primary interest \cite{farazmand2019extreme,pickering2022discovering,bollerslev1986generalized,engle2001garch}. Thus, structure inferred only from deterministic responses or conditional means can miss variables that are essential for representing the stochastic behavior of the target. 

\begin{figure}[t]
\centering
\includegraphics[width=0.7\linewidth]{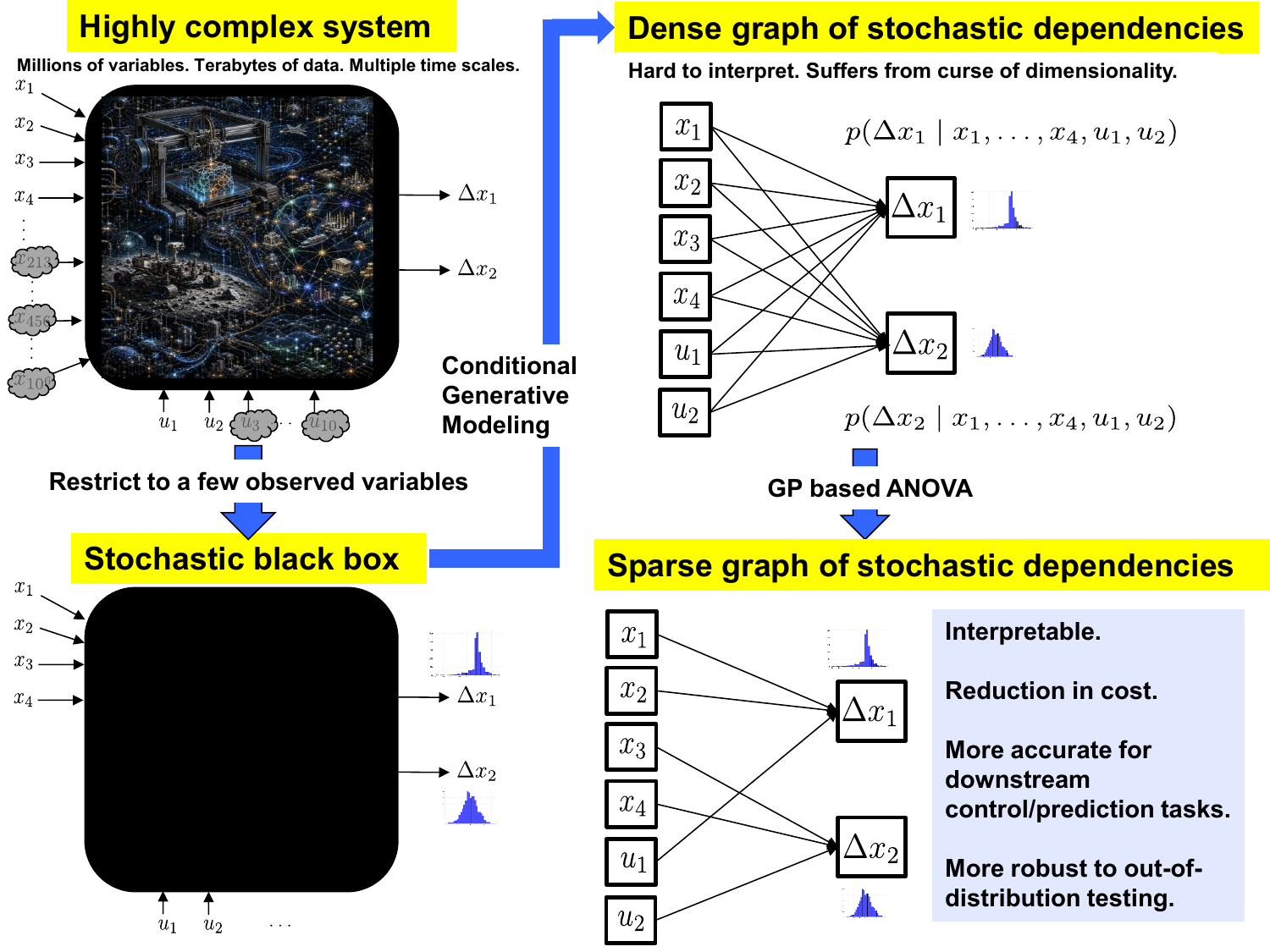}
\caption{
Structure discovery for sparse stochastic digital twins.
}
\label{fig:fig1}
\end{figure}

To address this problem, we characterize model structure directly at the level of the conditional law. Given a target quantity $y$ and candidate inputs $X=(x_1,\ldots,x_n)$, we represent the conditional distribution using a conditional generative model
\begin{equation}
    y\mid X \;\overset{d}{\approx}\; \mathcal{G}_\theta(X,z),
    \qquad z\sim p_0.
\end{equation}
Although the conditional generative model provides a flexible representation of the full conditional law, it does not directly expose how strongly individual input variables contribute to that law.
We therefore distill the learned generator $\mathcal{G}_\theta$ into a Gaussian-process (GP) representation and apply kernel mode decomposition (KMD) \cite{owhadi2021kernel} to obtain variable-wise nonlinear analysis of variance (ANOVA) contribution scores. These scores quantify how strongly the learned conditional law depends on individual variables, enabling comparison across variables, operating regimes, and time, while also providing a basis for constructing sparse stochastic surrogates. We iteratively and greedily remove variables with the smallest contribution scores and monitor the GP noise-to-signal ratio (NSR) \cite{bourdais2024codiscovering}; a sharp increase in NSR signals a loss of fidelity and provides a stopping criterion for the reduction. Figure~\ref{fig:fig1} summarizes the framework.

The discovered dependence structure has both scientific and operational value. Contribution scores provide an interpretable description of how individual variables enter the conditional law and can reveal changes in dependence across operating regimes or time. The same structure can also be used to construct sparse stochastic digital twins that reduce the dimension of data-driven modeling and downstream learning or control problems.

The contribution of this work is twofold. First, we show that higher model fidelity does not always lead to better downstream performance. In a modified Lunar Lander problem, an eight-dimensional sparse stochastic surrogate yields substantially better control performance than the exact 109-dimensional simulator under the same policy-learning setup, despite being less faithful to the transition dynamics. Second, we introduce a structure-discovery framework for stochastic digital twin modeling that applies both to task-oriented reduction of known high-dimensional models and to data-driven construction of sparse stochastic surrogates.
Our numerical studies validate the framework across several settings: a simple algebraic system demonstrates dependence hidden in the full conditional distribution; a 102-dimensional stochastic heat-equation problem shows that the method recovers sparse structure and improves a data-driven surrogate relative to a full high-dimensional surrogate; and an economic-data example shows that the contribution scores reveal time-varying memory structure in the dependence between geopolitical risk and oil-price dynamics.

\section{Results}\label{sec:reuslts}

\subsection{Algebraic system with hidden conditional dependence}

\begin{figure}[t]
    \centering
    \includegraphics[width=1.0\linewidth]{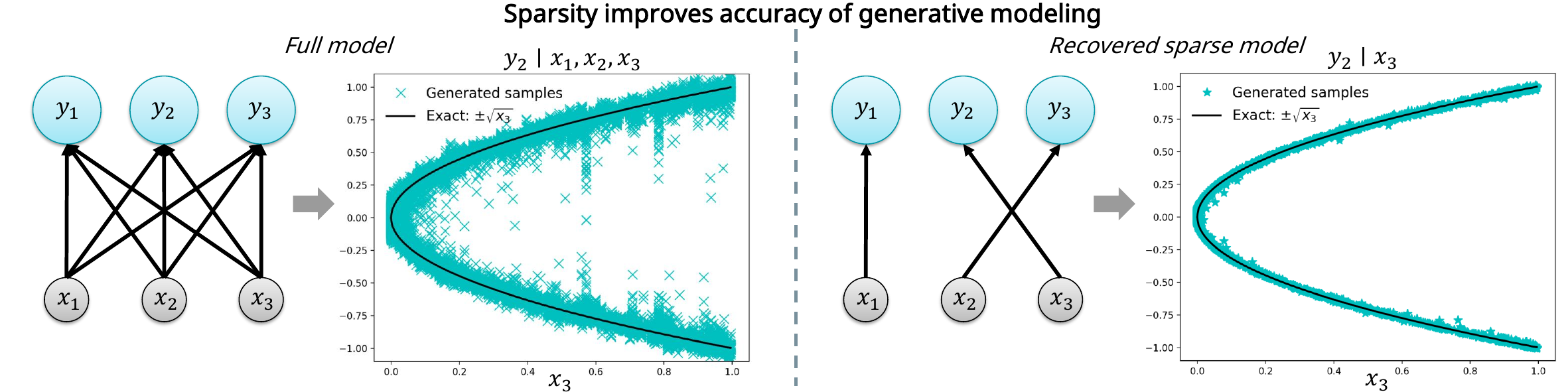}
    \caption{Hidden conditional dependence and sparse conditional generative
    modeling in the algebraic system.}
    \label{fig:algebraic_0}
\end{figure}

We first consider a simple algebraic system to illustrate two basic properties of the proposed framework: dependence can be hidden in the conditional distribution, and identifying sparse dependence structure can improve conditional generative modeling.
Specifically, let
\[
y_1 = x_1\omega, \qquad
y_2 = \varepsilon \sqrt{x_3}, \qquad
y_3 = \sqrt{1-\alpha(x_2)}\,\xi + \sqrt{\alpha(x_2)}\,\eta,
\]
where $\alpha(x)=\sin(\pi x)$, $x_3\ge 0$,
$\omega,\xi\sim \mathcal{N}(0,1)$, $\eta\sim\mathrm{Laplace}(0,1/\sqrt{2})$,
and $\varepsilon$ is an independent Rademacher random variable. In this setting,
\[
\mathbb{E}[y_1\mid x_1,x_2,x_3]
=
\mathbb{E}[y_2\mid x_1,x_2,x_3]
=
\mathbb{E}[y_3\mid x_1,x_2,x_3]
=0,
\qquad
\mathrm{Var}[y_3\mid x_1,x_2,x_3]=1.
\]
Hence, the dependence of $y_1$ on $x_1$, $y_2$ on $x_3$, and $y_3$ on $x_2$ is completely hidden from the conditional mean. The dependence of $y_3$ on $x_2$ is further hidden from the conditional variance and appears only through changes in the shape of the conditional distribution, as $x_2$ controls the relative contributions of the Gaussian and Laplace components.

Treating $\omega,\varepsilon,\xi,$ and $\eta$ as unobserved variables, the proposed method correctly recovers the sparse dependence structure: $\{x_1\}$ for $y_1$, $\{x_3\}$ for $y_2$, and $\{x_2\}$ for $y_3$. Moreover, as shown in Figure~\ref{fig:algebraic_0}, using the recovered sparse structure improves conditional generative modeling: for $y_2$, the model conditioned only on $x_3$ more accurately reproduces the exact conditional distribution than the full model conditioned on $(x_1,x_2,x_3)$. This simple example demonstrates both the need to characterize dependence at the level of the full conditional distribution and the benefit of removing unnecessary conditioning variables when constructing stochastic surrogates.

\subsection{Sparse model can outperform exact dynamics in downstream control}

Greater model fidelity does not necessarily imply better downstream performance when the additional fidelity substantially increases the dimension of the downstream learning problem. We illustrate this point using the \texttt{LunarLanderContinuous-v3} environment from Gymnasium \cite{towers2024gymnasium}. The physical state is eight-dimensional, consisting of the horizontal and vertical positions $(x,y)$, orientation $\theta$, linear velocities $(v_x,v_y)$, angular velocity $\omega$, and two leg-contact indicators $(l_l,l_r)$. The continuous controls $(a_1,a_2)$ govern the main engine and lateral booster. At each time step, the engines apply stochastic impulses to $(v_x,v_y,\omega)$, after which the remaining motion is updated deterministically. The control objective is to land the vehicle safely on the designated landing pad.

We modify this environment by introducing 101 additional state variables: 100 auxiliary variables $z_1,\ldots,z_{100}$ and a fuel variable $f$. The auxiliary variables are one-way coupled to the physical state-action dynamics: they depend on the physical state and controls but do not feed back into the lander dynamics. In contrast, the fuel variable weakly modulates the main-engine thrust and therefore has a nonzero but small effect on the stochastic velocity increments $(\Delta v_x,\Delta v_y,\Delta\omega)$. The resulting exact model has a 109-dimensional state.

\begin{figure}[ht!]
\centering\includegraphics[width=1.0\linewidth]{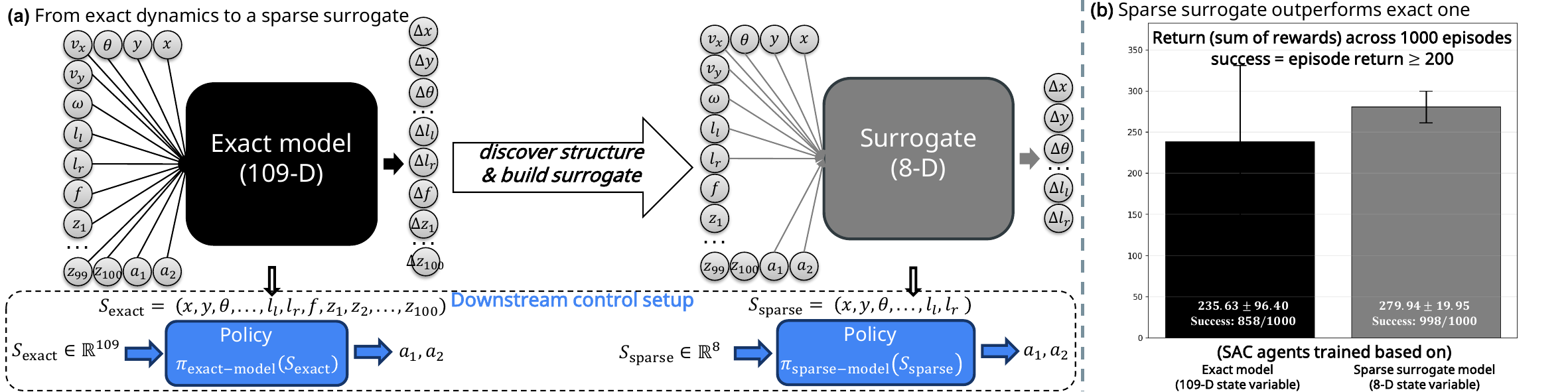}
    \caption{Model reduction in the modified Lunar Lander problem. (a): Our method identifies an eight-dimensional surrogate that retains the dominant thrust dependencies while removing the 100 auxiliary variables and the weak dependence on fuel. (b): SAC policies trained using the sparse surrogate and the exact high-dimensional simulator are evaluated in the same exact environment.}
    \label{fig:lunar_2}
\end{figure}

We apply the proposed method to the stochastic thrust dynamics, treating the 109 state variables and two controls as candidate inputs. The method removes the 100 auxiliary variables and weak fuel dependence, retaining only the thrust inputs $(\theta, a_1, a_2)$. Combining the reduced thrust model with the known Newtonian dynamics yields an eight-dimensional surrogate defined on the original physical state; see Figure~\ref{fig:lunar_2}(a). Additional Lunar Lander results and implementation details are provided in Appendix B.2. Because the true, though weak, dependence on fuel is discarded, the sparse surrogate is necessarily less faithful to the exact transition dynamics.

We then ask whether this loss of fidelity necessarily reduces downstream control performance. Using the same Soft Actor--Critic (SAC) \cite{haarnoja2018soft} architecture, hyperparameters, and training budget, we train policies either in the exact 109-dimensional simulator or in the reduced eight-dimensional surrogate, and evaluate both policies in the same exact environment. As shown in Figure~\ref{fig:lunar_2}(b), the policy trained with the reduced surrogate achieves a mean return of $279.94\pm19.95$ with $998/1000$ successful landings, whereas the policy trained with the exact model achieves $235.63\pm96.40$ with $858/1000$ successful landings. Thus, despite sacrificing transition fidelity, the sparse surrogate yields better downstream control by substantially reducing the dimensionality of policy learning.

\subsection{Sparse surrogates improve high-dimensional data-driven digital twins}

\begin{figure}[ht!]
\centering\includegraphics[width=1.0\linewidth]{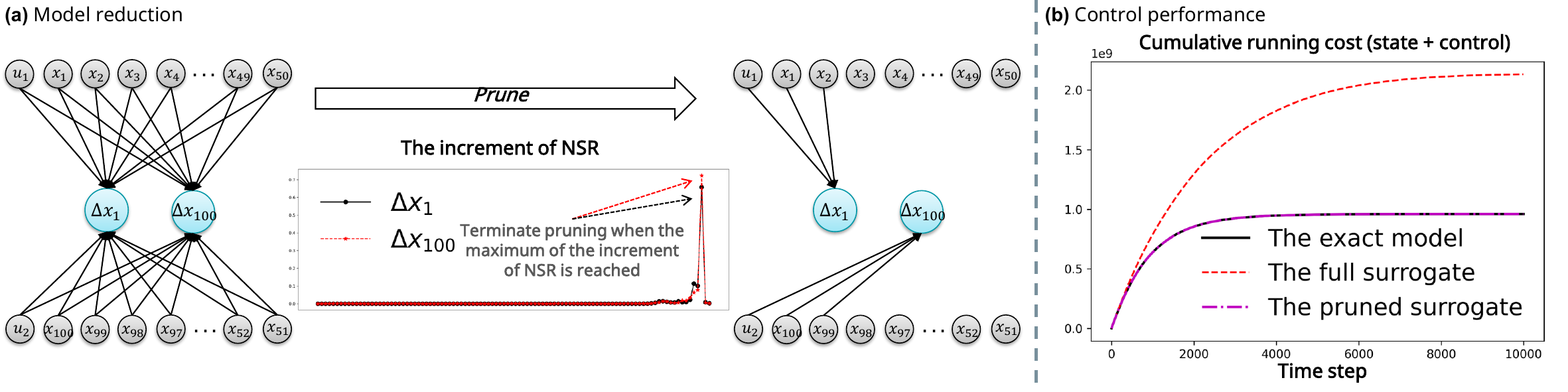}
    \caption{High-dimensional structure in stochastic heat equation with boundary control. (a): Model reduction for the two learned boundary increments. (b): Control performance comparison.
    }
    \label{fig:heat}
\end{figure}

We next consider a high-dimensional data-driven digital twin for a one-dimensional stochastic heat equation with Neumann boundary control, given by $\partial_t T=\kappa \partial_{rr}T+\sigma T\xi$ on $r\in(0,1)$, with boundary conditions $-\kappa \partial_rT(0,t)=u_1(t)$ and
$\kappa \partial_rT(1,t)=u_2(t)$. After spatial discretization using $m=100$ nodes, let $x_i(t_n)\approx T(r_i,t_n), i=1,...,m$.
We learn the stochastic updates of the two controlled boundary nodes, while retaining the known finite-difference updates for the interior nodes. The boundary increments take the form
$$\Delta x_1 = -2b x_1 + 2b x_2 + c u_1 + x_1\omega_1, \qquad \Delta x_{100} = 2b x_{99} - 2b x_{100} + c u_2 + x_{100}\omega_{100},$$
where $b$ and $c$ are determined by the discretization and $\omega_i$ denotes the discretized multiplicative noise. The true dependence structure is therefore sparse: $\Delta x_1$ depends only on $\{x_1,x_2,u_1\}$ and $\Delta x_{100}$ only on $\{x_{99},x_{100},u_2\}$. Starting from the full candidate set $\{x_1,\ldots,x_{100},u_1,u_2\}$, the proposed method recovers exactly these structures; see Figure~\ref{fig:heat}(a). The NSR remains nearly unchanged while irrelevant inputs are removed and increases sharply once pruning begins to remove an essential input, providing a clear stopping criterion in the 102-dimensional candidate space. Implementation and MPC details are provided in Appendix B.3.

We then examine whether exploiting this structure improves data-driven digital twin modeling. We train two stochastic surrogates for the boundary updates: a full surrogate conditioned on all 102 candidate inputs and a sparse surrogate conditioned only on the variables identified by the proposed method. Both surrogates are embedded in the same stochastic model predictive control (MPC) framework, with identical interior dynamics and control objective. Across ten independent trials, MPC using the sparse surrogate achieves running costs close to those obtained with the known dynamics, whereas MPC using the full surrogate yields substantially larger cost; see Figure~\ref{fig:heat}(b). Thus, in a high-dimensional data-driven setting, the discovered sparse dependence structure yields a more effective stochastic surrogate than one conditioned on all candidate inputs.

\subsection{Dependence analysis reveals changing memory structure in economic data} 

\begin{figure}[t!]
    \centering
    \includegraphics[width=1.0\linewidth]{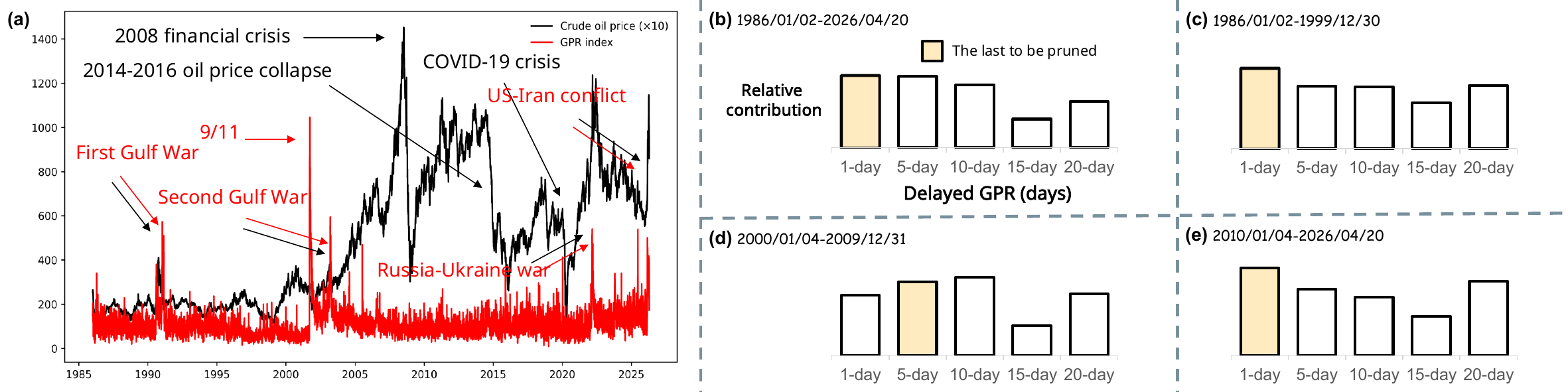}
    \caption{
    Delayed dependence analysis in economic markets. (a): Aligned WTI crude oil price and GPR index data. (b)--(e): First-iteration relative contribution scores of delayed GPR variables across time periods; the highlighted bar denotes the variable pruned last.  
    }
    \label{fig:economic}
\end{figure}

Lastly, we demonstrate the proposed method as an analytical tool for real economic data. 
We treat the conditional transition kernel of oil returns as a data-driven stochastic digital twin and study how its dependence on geopolitical risk changes across market regimes. 
We use the daily West Texas Intermediate (WTI) crude oil prices from the Federal Reserve Bank of St.\ Louis and the daily geopolitical risk (GPR) index~\cite{caldara2022measuring}, covering the period from 1986/01/02 to 2026/04/20 (Figure~\ref{fig:economic}(a)).
Let $r_n$ denote the daily log-return of oil prices ($r_n = \log(\text{Oil}_{n+1}) - \log(\text{Oil}_{n})$). We model the transition kernel
\begin{equation}\label{eq:economic}
    p(r_{n}|r_{ n-1}, r_{n-5}, r_{n-10}, r_{n-15}, r_{n-20}, \log \text{GPR}_n, \log \text{GPR}_{n-4}, \log \text{GPR}_{n-9}, \log \text{GPR}_{n-14}, \log \text{GPR}_{n-19}).
\end{equation}
Because $r_n$ is defined over the interval from retained observation
$n$ to $n+1$, the GPR terms indexed by $n,n-4,n-9,n-14,n-19$ correspond to the 1-, 5-, 10-, 15-, and 20-observation
lags, respectively.
We apply the proposed method to the GPR lag variables to analyze how delayed geopolitical-risk information enters the oil-return transition kernel. 

The method retains all five GPR lags as conditioning inputs, so we further examine their relative contribution scores and pruning order.
Figure~\ref{fig:economic}(b)--(e) reports two diagnostics for the full 1986-2026 period and three sub-periods: the first-iteration relative contribution scores of the five GPR lags, and the lag pruned last by the iterative pruning procedure. Over the full period, the 1-day lag has the largest first-iteration contribution and is also the last lag to be pruned. In contrast, the dominant surviving lag changes across sub-periods: the 1-day lag is last pruned in 1986-1999 and 2010-2026, whereas the 5-day lag is last pruned in 2000-2009. These changes indicate that the delayed dependence structure between geopolitical risk and oil returns is not stationary across time. The 2000-2009 period coincides with major geopolitical and economic disruptions, including 9/11 and the 2008 financial crisis, and the detected dependence structure in this period shifts toward a longer lag.
Eight shorter sub-periods are also analyzed with the proposed method, revealing a more detailed evolution of the delayed dependence structure; see Appendix B.4 for the results.
These results illustrate how the proposed method can be used not only to construct sparse stochastic digital twins, but also to analyze their dependence structure and reveal changes in memory from observational data.

\subsection{Additional validation across stochastic systems}

Beyond the main examples above, we further evaluate the proposed framework on a broader collection of stochastic systems, including nonlinear stochastic differential equations, chemical reaction networks, multiscale dynamics, coarse-grained systems with memory, partially observed control systems, and high-dimensional robotic dynamics. These additional examples test the robustness and generality of the dependence-discovery framework across different forms of stochasticity, dimensionality, and model structure. Detailed results are provided in Appendix C.

\section{Discussion}\label{sec:conclusions}

We show that model fidelity and downstream utility are not always aligned. Retaining additional state variables can improve fidelity while making learning, optimization, or control more difficult. In the modified Lunar Lander example, the sparse eight-dimensional surrogate is less faithful than the exact 109-dimensional model, yet it yields better downstream control under the same policy-learning setup.

We propose a framework for discovering such structure directly at the level of the conditional law. The framework can be used both to construct sparse stochastic surrogates from data, as in the stochastic heat-equation example, and to reduce known high-dimensional models for downstream tasks, as in Lunar Lander. The same dependence scores can also support interpretation, as illustrated by the changing memory structure identified in the economic data.

Several limitations remain. The framework depends on accurate conditional generative modeling, and the NSR stopping criterion can be less clear when dependence strengths vary gradually. More generally, this work shows that sparse stochastic twins can outperform higher-fidelity models in some downstream settings, but does not yet characterize when this tradeoff will favor sparsity. Developing such a characterization is an important direction for future work.

\section{Methods}

After training the conditional generative model $\mathcal{G}_\theta$ (details in Appendix A), we generate synthetic samples from the learned conditional law. For inputs
$X^i=(x_1^i,\ldots,x_n^i)$ and reference samples $z^i\sim p_0$, we compute
$\hat y^i=\mathcal{G}_\theta(X^i,z^i)$, yielding a dataset for distillation 
$\{(X^i,z^i,\hat y^i)\}_{i=1}^M$ that we use to train our GP surrogate
$f(X,z)\approx \mathcal{G}_\theta(X,z)$. 

For a kernel $K_s$ and regularization parameter $\gamma$, the surrogate is
obtained by kernel ridge regression. That is, 
\begin{equation}\label{eq:GP_ridge}
    f
    =
    \arg\min_{\varphi\in\mathcal{H}_{K_s}}
    \|\varphi\|_{K_s}^2
    +
    \frac{1}{\gamma}
    \sum_{i=1}^M
    |\varphi(X^i,z^i)-\hat y^i|^2,
\end{equation}
where $\mathcal{H}_{K_s}$ denotes the reproducing kernel Hilbert space (RKHS)
associated with $K_s$. Here we use the fact that sums of kernels induce corresponding
minimum-norm decompositions of RKHS functions \cite{owhadi2021kernel}. Namely,
if $k=k_a+k_b$ and $f_k\in\mathcal{H}_k$, then its kernel-mode components are
defined by
\[
(f_a,f_b)
=
\arg\min_{g_a,g_b}
\left\{
\|g_a\|_{k_a}^2+\|g_b\|_{k_b}^2
:\;
g_a+g_b=f_k,\;
g_a\in\mathcal{H}_{k_a},\;
g_b\in\mathcal{H}_{k_b}
\right\}.
\]
They satisfy $f_k=f_a+f_b$ and
$\|f_k\|_k^2=\|f_a\|_{k_a}^2+\|f_b\|_{k_b}^2$, so
$\|f_a\|_{k_a}^2/\|f_k\|_k^2$ measures the fraction of the RKHS signal energy that is
attributed to the component associated with $k_a$.
We use this decomposition to quantify variable-wise contribution scores.

Problem \eqref{eq:GP_ridge} can be interpreted as regression with the augmented kernel
$K_s+\gamma\delta$, where $\delta$ is the white-noise kernel. The corresponding
decomposition separates the signal component associated with $K_s$ from the
noise component associated with $\gamma\delta$. We define
\begin{equation}
    \mathrm{NSR}:=\frac{V_n}{V_s+V_n},
    \qquad
    V_s:=\|f\|_{K_s}^2,
    \qquad
    V_n:=\|f_{\gamma\delta}\|_{\gamma\delta}^2=
    \frac{1}{\gamma}
    \sum_{i=1}^M |f(X^i,z^i)-\hat y^i|^2 .
\end{equation}
A large increase in NSR during pruning indicates that the removed variable makes a substantial contribution to the learned conditional law.
The second decomposition quantifies the contribution of individual variables to the learned conditional law. 
To identify variables, we use the product kernel
\begin{equation}
    K_s((X,z),(X',z'))
    =
    (1+k_z(z,z'))
    \prod_{i=1}^n (1+k_x(x_i,x_i')),
\end{equation}
where $k_x$ and $k_z$ are universal kernels, such as radial basis function (RBF) or Mat\'ern kernels.
Expanding this product yields a sum of interaction kernels over subsets of the
covariates, inducing a nonlinear ANOVA decomposition
\cite{owhadi2021kernel, bourdais2024codiscovering}. For each variable $x_j$, we set
\[
    K_j=(1+k_z(z,z'))\,k_x(x_j,x_j')
    \prod_{i\neq j}(1+k_x(x_i,x_i')),
    \qquad
    K_{-j}=K_s-K_j.
\]
The kernel $K_j$ collects all ANOVA components involving $x_j$, while $K_{-j}$
collects the components that do not involve $x_j$. Applying the minimum-norm
splitting to $K_s=K_j+K_{-j}$ gives $f=f_j+f_{-j}$, and we define
\[
    c_j:=\frac{\|f_j\|_{K_j}^2}{\|f\|_{K_s}^2}\in[0,1].
\]
The score $c_j$ measures the fraction of RKHS signal energy, or equivalently the
GP/nonlinear-ANOVA variance contribution, attributable to components involving
$x_j$ \cite{owhadi2021kernel,bourdais2024codiscovering}; variables with smaller scores therefore make weaker contributions to the learned conditional law.

Model reduction proceeds by iterative pruning. At each step, the variable
with the smallest contribution score is removed and the GP surrogate is retrained
on the reduced input set. The pruning sequence is run to completion, the NSR is
recorded at each step, and the selected set of input variables is the set of variables remaining immediately before the largest NSR increment. 

Unlike conditional independence tests \cite{zhang2012kernel,gretton2007kernel},
which return binary accept/reject decisions, this procedure assigns each
candidate variable a continuous contribution score through a variance
decomposition of the learned conditional generative model. This supports pruning when
finite data or indirect associations lead to weak contributions from some
variables. It also avoids testing variables against large conditioning sets, whose cost and reliability degrade with dimension \cite{zhang2012kernel}. Instead, a sharp increase in NSR indicates that pruning has removed an input that is important for representing the learned conditional law.

\subsection*{Acknowledgments}
We acknowledge support from the Air Force Office of Scientific Research under MURI award number FOA-AFRL-AFOSR-2023-0004 (Mathematics of Digital Twins), the Department of Energy under award number DE-SC0023163 (SEA-CROGS: Scalable, Efficient, and Accelerated Causal Reasoning Operators, Graphs and Spikes for Earth and Embedded Systems), the National Science Foundation under award number 2425909 (Discovering the Law of Stress Transfer and Earthquake Dynamics in a Fault Network using a Computational Graph Discovery Approach) and the Vannevar Bush Faculty Fellowship under ONR-N000142512035.
This material is also based upon work  supported by the Defense Advanced Research Projects Agency (DARPA) under Agreement No. HR00112590112. Approved for public release; distribution is unlimited.

\subsection*{Data availability}

Data for the numerical examples are generated using the simulation code provided in the accompanying code repository. The West Texas Intermediate (WTI) crude-oil price data used in the economic example are publicly available from the Federal Reserve Bank of St. Louis. The geopolitical risk (GPR) index data are publicly available from the website of Caldara and Iacoviello at \url{https://www.matteoiacoviello.com/gpr.htm}. The preprocessed economic data used in this study are provided in the accompanying code repository.

\subsection*{Code availability}
The code used in this study is available at \url{https://github.com/ZongrenZou/sparse-digital-twins}.

\bibliographystyle{plain} 
\bibliography{references}

\newpage
\appendix
\section{Conditional generative modeling via flow matching}
\label{sec:flow_matching}
In this section, we describe the conditional generative modeling stage. The
objective is to learn a generative representation for the conditional law of $y$ given
$X=(x_1,\ldots,x_n)$. We use flow matching models \cite{lipman2022flow}, which
learn a velocity field transporting a simple base distribution to the target
conditional distribution through the ordinary differential equation
\begin{subequations}\label{eq:ode}
    \begin{align}
        \frac{d\zeta_t}{dt} &= v_\theta(t,\zeta_t,X), \qquad t\in(0,1),\\
        \zeta_0 &= z, \qquad z\sim p_0,
    \end{align}
\end{subequations}
where $v_\theta$ is a neural network parameterized by $\theta$, and $p_0$ is a
tractable base distribution, typically a standard Gaussian. The corresponding
solution map defines the conditional generative model
\begin{equation}
    \hat y = \mathcal{G}_\theta(X,z) := \zeta_1,
\end{equation}
where $\zeta_1$ is obtained by integrating \eqref{eq:ode} from $t=0$ to $t=1$.
Training is performed by the flow matching regression objective. Given data
pairs $(X,y)$ and an independent sample $z\sim p_0$, define the linear
interpolation
\begin{equation}
    \tilde\zeta_t = (1-t)z + ty,
\end{equation}
whose target velocity is $\dot{\tilde\zeta}_t = y-z$. We then learn $v_\theta$
by solving
\begin{equation}\label{eq:optimization}
    \min_{\theta}\;
    \mathbb{E}_{(X,y),\,z\sim p_0,\, t\sim \mathcal{U}(0,1)}
    \left[
    \left\|(y-z)-v_\theta(t,\tilde\zeta_t,X)\right\|^2
    \right].
\end{equation}
After training, samples from the learned approximation of $y\mid X$ are
generated by drawing $z\sim p_0$ and evaluating
$\hat y=\mathcal{G}_\theta(X,z)$. This procedure is known to converge in the limit of infinite data \cite{albergo_stochastic_2023} under sufficient regularity and well-specified approximation class assumptions.

\section{Additional results and implementation details}\label{sec:appendix:additional_details}
We provide additional results and implementation details for the numerical examples.
In all numerical examples, unless stated otherwise, we use (1) the Adam optimizer \cite{kingma2014adam} with learning rate 1e-4 to train the flow matching model, (2) the Diffrax ODE solver \cite{kidger2021on} (adaptive fifth-order Runge--Kutta integration with PID step-size control; 1e-5 relative and absolute tolerances) to generate samples from the trained flow matching model, and (3) the nugget $\gamma = 1.0$ in Gaussian process (GP) regression, where $\gamma$ is the regularization parameter in \eqref{eq:GP_ridge}.

\subsection{Algebraic system with hidden conditional dependence}

\begin{figure}[ht!]
    \centering
    \includegraphics[width=0.5\linewidth]{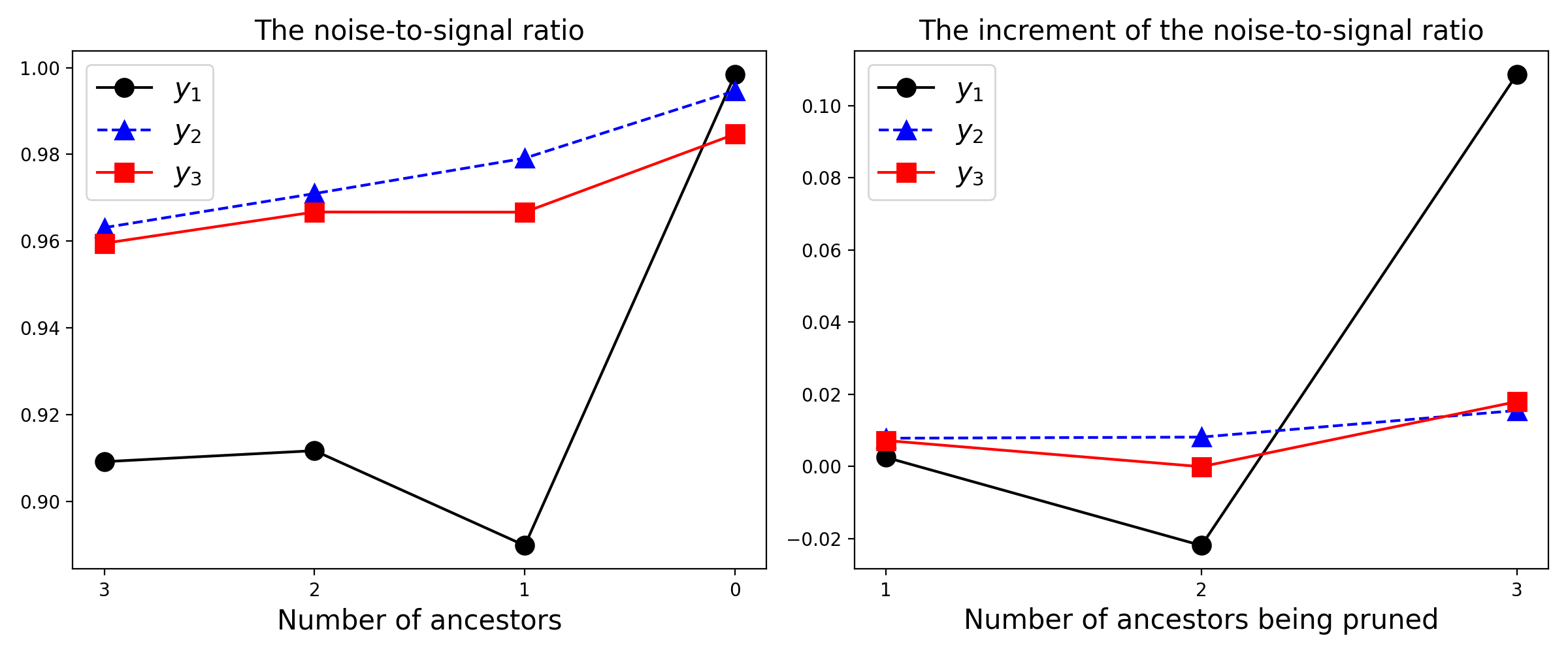}
    \caption{Noise-to-signal ratio during iterative variable removal for the algebraic system with hidden conditional dependence.}
    \label{fig:algebraic}
\end{figure}

We use $1{,}000$ data points for dependence discovery for $y_1$ and $y_2$, and $10{,}000$ data points for $y_3$. 
More data points are required for accurately discovering the dependence structure of $y_3$ than for $y_1$ and $y_2$, because the dependence appears only through the shape of the conditional distribution rather than through the mean or variance, as discussed in the main text.
In particular, data of $x_1$ and $x_2$ are sampled from $\mathrm{Unif}([-0.5, 0.5])$ and $\mathrm{Unif}([0, 1])$, respectively, and data $y_1$ and $y_3$ are generated from the exact formula using sampled inputs and random variables with $\alpha(x):=\sin(\pi x)$. For $y_2$, we sample $y_2\sim\mathrm{Unif}([-1,1])$ and set $x_3=y_2^2$.

For dependence discovery, we model the distributions of $y_1, y_2, y_3$ individually by applying the proposed method independently to each output. For each $y_k, k=1,2,3$, we use a fully connected neural network (FNN) with four hidden layers of 64 neurons each and rectified linear unit (ReLU) activation function to parameterize the velocity field, with batch size and number of epochs set to $1{,}000$ and $5{,}000$, respectively. In the second stage, we use $5{,}000$ distilled samples and the RBF kernel with lengthscale $5.0$ for $x_k, k=1,2,3$. 
For $y_1$ and $y_3$, we use lengthscale $1.0$ for the random variable $z$ and nugget $0.1$, while for $y_2$, we use lengthscale $0.1$ for $z$ and nugget $1.0$. This choice is made based on the observation that the conditional distribution of $y_2\mid x_1, x_2, x_3$ is a two-point distribution concentrated at $\pm\sqrt{x_3}$. This discrete structure is poorly suited to both flow matching and GP regression, motivating the adjusted kernel hyperparameters.
Figure~\ref{fig:algebraic} shows the corresponding noise-to-signal ratio (NSR) trajectories during iterative variable removal. The NSR remains nearly unchanged while nonessential variables are removed and increases sharply once an essential variable is discarded; the largest NSR increment therefore determines the stopping point.

\subsection{The Lunar Lander problem}\label{sec:appendix:lunar}

We provide additional results and implementation details for the two Lunar Lander experiments. The position-distribution-shift experiment is reported here in full, while the high-dimensional model-reduction experiment is presented in the main text and supplemented here with implementation details.

\subsubsection{Sparse stochastic surrogate enables transfer under position-distribution shift}\label{sec:appendix:lunar_case_a}

\begin{figure}[ht!]
\centering\includegraphics[width=1.0\linewidth]{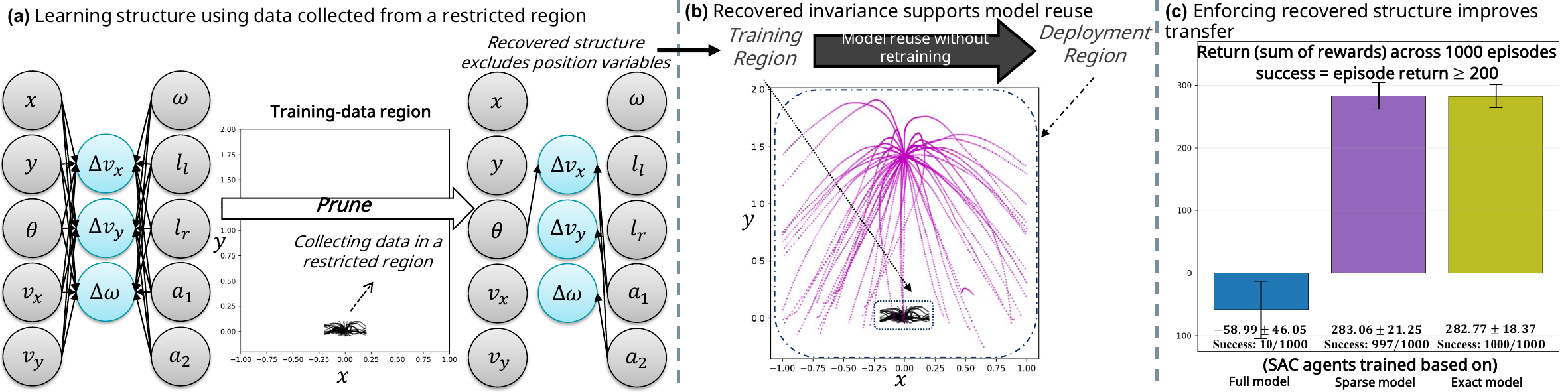}
    \caption{Lunar Lander transfer under position-distribution shift. (a): Structure learned from data collected in a restricted region; the recovered thrust structure excludes the position variables $(x,y)$. (b): The resulting position invariance provides a structural basis for reusing the learned model in a different deployment region. (c): Downstream SAC performance for policies trained with the full surrogate, sparse surrogate, and true dynamics.
    }
    \label{fig:lunar}
\end{figure}

One crucial challenge in digital twin modeling is that the distribution under which data are collected often differs from the distribution encountered during deployment. A learned dynamics model may therefore be trained using data from only a restricted region of the state space and later be used under a shifted operating distribution. Structural information can help determine which components of the learned model should remain valid under such shifts.

We illustrate this idea using the stochastic thrust dynamics of the Lunar Lander problem \cite{towers2024gymnasium}. The state contains the horizontal and vertical positions $(x,y)$, orientation $\theta$, linear velocities $(v_x,v_y)$, angular velocity $\omega$, and the two leg-contact indicators $(l_l,l_r)$, while the controls are the main-engine and side-engine inputs $(a_1,a_2)$. We focus on the stochastic thrust stage, where the engines apply stochastic impulses to the velocity variables. 
Training data are collected from trajectories in a restricted position regime, generated by initializing the lander near $y=0$ and terminating each trajectory once it leaves $|x|<0.2, |y|<0.1$. The candidate inputs include the full state-control vector.
Applying the proposed dependence analysis to these data yields a sparse thrust structure in which the position variables $x$ and $y$ are removed; see Figure~\ref{fig:lunar}(a). The recovered sparse dependence structure therefore indicates that the learned thrust mechanism is invariant to position. This provides a structural basis for reusing the learned thrust model when the deployment distribution shifts in $(x,y)$, without retraining the model solely because the lander operates in a different region; Figure~\ref{fig:lunar}(b) illustrates this setting.

Importantly, identifying this invariance is not sufficient if the learned surrogate is still allowed to depend on $(x, y)$. With finite training data, a full surrogate conditioned on all candidate variables can acquire spurious position dependence that is harmless inside the training region but becomes detrimental after the position distribution shifts. In contrast, the sparse surrogate excludes $(x, y)$ by construction and therefore enforces the discovered invariance.
To evaluate this effect, we train full and sparse thrust surrogates using the same restricted-region data, replace the thrust stage with each surrogate model to train a Soft Actor--Critic (SAC) policy, and then test the resulting policies in the true environment under the shifted position distribution. As shown in Figure~\ref{fig:lunar}(c), the policy trained with the sparse surrogate achieves performance comparable to the policy trained in the true environment, with $997$ successful landings out of $1{,}000$ episodes and mean return $283.06\pm21.25$. In contrast, the policy trained with the full surrogate succeeds in only $10$ of $1{,}000$ episodes and has a mean return of $-58.99\pm46.05$. Thus, the discovered structure serves not only as an interpretable description of the thrust dynamics, but also as actionable information for constructing a surrogate that can be reliably reused under position-distribution shift.

\subsubsection*{\textbf{Implementation details}}

We consider the \texttt{LunarLanderContinuous-v3} environment from Gymnasium \cite{towers2024gymnasium} (version 1.2.3). The continuous action $a=(a_1, a_2) \in[-1, 1]^2$ controls the main engine throttle and lateral booster throttle, respectively.

The dynamics decompose into two stages at each time step: (1) a thrust stage, in which the engines apply instantaneous velocity impulses to the lander, perturbing ($v_x$, $v_y$, $\omega$), and (2) a Newtonian stage, which integrates the equations of motion, updating positions and angles and handling contact dynamics, deterministically given the post-impulse velocities. We use the default environment configuration: wind and turbulence are disabled, so the only stochasticity enters through the engine impulse dispersion. The terrain is procedurally generated at each episode reset, and the helipad is centered horizontally.
Under the default initialization, the lander begins near the top center of the viewport with small random perturbations to position, velocity, and angle. Experiment-specific modifications to the initialization are described below; the reward function, termination conditions, and observation normalization are unchanged.

Training data are collected under a uniform random policy with actions drawn independently as $a\sim \mathrm{Unif}([-1, 1]^2)$ at each step. We modify the lander's initial vertical coordinate in the simulator to 4.0 in the unnormalized environment coordinates, which places the normalized observed vertical position near $y=0$. We then simulate $50$ independent episodes of at most $1{,}000$ steps, terminating early upon crash, successful landing, or when the lander leaves the restricted position region $\{|x|<0.2, |y|<0.1\}$. The velocity increments $\Delta v_x, \Delta v_y, \Delta \omega$ are computed as the difference in the lander's linear and angular velocities immediately before and after the engine impulses are applied, prior to the Newtonian integration step. Each transition contributes one record to the dataset, consisting of a 10-dimensional input vector
\begin{equation}\nonumber
    X = [x, y, v_x, v_y, \theta, \omega, l_l, l_r, a_1, a_2]
\end{equation}
concatenating the full 8-dimensional state observed before the step with the 2-dimensional action, together with the three scalar thrust-stage outputs $\Delta v_x, \Delta v_y, \Delta\omega$.

For dependence discovery, we model the conditional law of $\Delta v_x, \Delta v_y, \Delta\omega$ jointly. The velocity field is parameterized by an FNN with four hidden layers of 64 neurons and sigmoid linear unit (SiLU) activation function, trained with batch size $1{,}000$ for $10{,}000$ epochs. Structure discovery uses $10{,}000$ distilled samples and an RBF kernel with lengthscale $5.0$ for all state, control, and latent variables.

For the downstream control experiment, we restore the default Lunar Lander initialization used by Gymnasium, so the lander begins near the top center of the viewport rather than near the restricted region used for training-data collection. Thus, the thrust surrogates are trained only on data collected from the localized region $|x|<0.2$, $|y|<0.1$, but are subsequently used to train control policies in an environment whose state distribution spans a substantially different range of positions.

The control objective is to land the lander softly on the helipad. The step reward is
\begin{equation}\nonumber
    r_n = \Delta S_n - 0.30\, m_n - 0.03\, s_n,
\end{equation}
where $\Delta S_n = S_n - S_{n-1}$ is the change in the shaping potential
\begin{equation}\nonumber
    S_n = -100\sqrt{x_n^2 + y_n^2} - 100\sqrt{v_{x,n}^2 + v_{y,n}^2} - 100|\theta_n| + 10\, l_{l,n} + 10\, l_{r,n},
\end{equation}
$m_n\in[0, 1]$ and $s_n\in[0, 1]$ are the main and side engine throttle magnitudes at step $n$ (nonlinear functions of $a_{1, n}$ and $a_{2, n}$, respectively), and $l_{l, n}, l_{r, n} \in \{0,1\}$ indicate leg ground contact. A terminal bonus of $+100$ is awarded when the lander comes to rest on the ground (the rigid body becomes inactive), and $-100$ when the lander hull contacts the ground or exits the horizontal viewport bounds. An episode is considered successful if the cumulative return is greater than or equal to $200$. We refer to \cite{towers2024gymnasium} for the full specification.

We evaluate the discovered structure by training a control policy in the surrogate environment and transferring it to the true environment without further training. The thrust stage is replaced by flow matching surrogates conditioned on the sparse dependence structure:
\begin{align}\nonumber
(\Delta v_x, \Delta v_y, \Delta\omega) = G(\theta, a_1, a_2, z),
\end{align}
where $z\sim\mathcal{N}(0, I_3)$ is a three-dimensional latent vector. The Newtonian stage remains identical to that of the true environment.
Policies are trained with SAC implemented in Stable-Baselines3~\cite{stable-baselines3}, using the tuned hyperparameters from RL Baselines3 Zoo~\cite{rl-zoo3} (version 2.8.0) for \texttt{LunarLanderContinuous-v3}. 
For the surrogate-based regimes, policy updates use only interactions with the corresponding surrogate environment. We use the same SAC training protocol across three regimes: (1) the true environment, (2) the full surrogate, and (3) the sparse surrogate, so that differences in downstream performance reflect the transition model used for policy training.

\subsubsection{Sparse surrogate can improve control even relative to exact dynamics: implementation details}\label{sec:appendix:lunar_case_b}

The original eight-dimensional Lunar Lander state is augmented by $100$ auxiliary variables $z_1,\ldots,z_{100}$ and a scalar fuel variable $f$, giving
\begin{equation}\nonumber
s^{\mathrm{exact}}
=
[x,y,v_x,v_y,\theta,\omega,l_l,l_r,z_1,\ldots,z_{100},f]
\in\mathbb{R}^{109}.
\end{equation}
The auxiliary variables $z_1,\ldots,z_{100}$ are one-way coupled to the physical state and control.
At each transition, the physical Lunar Lander state is first advanced from step $n$ to step $n+1$. The auxiliary variables are then updated according to 
\begin{equation}\nonumber
z_{j,n+1} = z_{j,n} + \Delta t\left(-\frac{z_{j,n}}{\tau}+d_{j,n}\right) + \sigma_z\sqrt{\Delta t}\,\xi_{j,n},
\end{equation}
where $\Delta t$ is the Lunar Lander simulation time step, $\tau=1$, $\sigma_z=0.05$, and $\xi_{j,n}\sim\mathcal N(0,1)$. The driving variables $d_{j,n}$ are assigned cyclically from the ten-dimensional physical state-action vector $$d_n^{\mathrm{base}} = [x_{n+1},y_{n+1},v_{x,n+1},v_{y,n+1},\theta_{n+1},\omega_{n+1},l_{l,n+1},l_{r,n+1},a_{1,n},a_{2,n}],$$
so that $d_{j,n} = d^{\mathrm{base}}_{((j-1)\bmod 10)+1,n}, j=1,\ldots,100$. Thus, the $100$ auxiliary variables are driven repeatedly by the ten physical state-action variables, but do not feed back into the physical Lunar Lander dynamics.

The fuel variable is initialized as $f_0=1$ and decreases with main-engine usage according to
\begin{equation}\nonumber
f_{n+1}
=
\operatorname{clip}\left(
f_n-\frac{m_n\Delta t}{T_f},0,1
\right),
\end{equation}
where $m_n$ is the main-engine throttle magnitude and $T_f=12$ is the full-throttle fuel-depletion time scale. Fuel weakly modulates the main-engine thrust by replacing the nominal engine power $P_{\rm main}$ with
\begin{equation}\nonumber
P_{\rm main}^{(n)}
=
P_{\rm main}\left[1-\alpha_f(1-f_n)\right],
\qquad \alpha_f=0.15.
\end{equation}
Thus, unlike the auxiliary variables, $f$ has a nonzero but deliberately weak effect on the physical transition.

For data generation, we use the default Lunar Lander initialization and do not modify the initial position. Training data are generated under a uniform random policy with $a_n\sim\mathrm{Unif}([-1,1]^2)$. We simulate $500$ independent episodes of at most $1{,}000$ steps, using the default termination conditions. At the beginning of each episode, $z_{j,0}\sim\mathcal N(0,1)$ independently and $f_0=1$. As in the previous experiment, the learning targets are the thrust-stage increments $(\Delta v_x,\Delta v_y,\Delta\omega)$ measured immediately before and after the engine impulses, prior to the Newtonian integration step. Each sample therefore uses the $111$-dimensional candidate input
\begin{equation}\nonumber
X =
[x,y,v_x,v_y,\theta,\omega,l_l,l_r,z_1,\ldots,z_{100},f,a_1,a_2],
\end{equation}
consisting of the $109$-dimensional exact state and the two-dimensional action.

We model the conditional law of $(\Delta v_x,\Delta v_y,\Delta\omega)$ jointly. The velocity field is parameterized by an FNN with four hidden layers of 64 neurons and SiLU activation function, trained with batch size $1{,}000$ for $2{,}000$ epochs. Structure discovery uses $5{,}000$ distilled samples and an RBF kernel with lengthscale $2.0$ for the state-control variables and $1.0$ for the latent variables. The nugget for GP regression is set to $5.0$.

The proposed dependence analysis removes $x,y,v_x,v_y,\omega,l_l,l_r,z_1,\ldots,z_{100},f$ from the thrust stage. The Newtonian stage evolves only the original eight Lunar Lander state variables $(x,y,v_x,v_y,\theta,\omega,l_l,l_r)$, while neither the Newtonian stage nor the reward function, which is identical to that used in the preceding position-distribution-shift experiment, depends on $z_1,\ldots,z_{100},f$. Therefore, removing $z_1,\ldots,z_{100},f$ from the thrust stage implies that these variables can be removed from the state representation used for downstream control. This yields an eight-dimensional surrogate state
\begin{equation}\nonumber
s^{\mathrm{sparse}}
=
[x,y,v_x,v_y,\theta,\omega,l_l,l_r]
\in\mathbb{R}^{8},
\end{equation}
and allows us to train a substantially lower-dimensional policy network. Because the sparse surrogate discards the true but weak dependence on fuel, it is less faithful to the exact transition dynamics. In contrast, training directly with the exact model requires the policy to take the full state $s^{\mathrm{exact}}\in\mathbb{R}^{109}$ as input. The SAC training protocol is the same as in the preceding experiment.

\subsection{The stochastic heat equation boundary control problem}\label{sec:appendix:heat}

We set $\sigma=0.01$ and $\kappa=0.01$. The spatial domain $[0,1]$ is discretized into $m=100$ equispaced grid points $r_i = (i-1)h$, $h = 1/(m-1)$, $i = 1, \ldots, m$. Applying second-order finite differences with ghost point elimination for the Neumann boundary conditions, and forward Euler time integration with step size $\Delta t=0.005$, yields the discrete-time stochastic system
\begin{equation}
    X_{n+1} = A X_n + B U_n + \text{diag}(X_n)\omega_n, \qquad \omega_n \sim \mathcal{N}(0, \sigma^2 I_m),
    \label{eq:heat_discrete}
\end{equation}
where $X_n=(x_{1, n}, ..., x_{m,n})^\top\in \mathbb{R}^m$ is the discretized temperature field at time step $n$ and $U_n = (u_{1, n}, u_{2, n})^\top\in \mathbb{R}^2$ is the control imposed on the boundaries. The state transition matrix $A \in \mathbb{R}^{m \times m}$ is tridiagonal and sparse, given by
\begin{equation}\nonumber
    A = I_m + r \tilde{A}, \qquad r = \frac{\kappa \Delta t}{h^2},
\end{equation}
where $\tilde{A}$ is the discrete Laplacian with Neumann boundary conditions:
\begin{equation}\nonumber
    \tilde{A} = \begin{pmatrix}
        -2 & 2  &        &        & \\
         1 & -2 & 1      &        & \\
           & \ddots & \ddots & \ddots & \\
           &        & 1      & -2    & 1 \\
           &        &        & 2     & -2
    \end{pmatrix}.
\end{equation}
The control input matrix $B \in \mathbb{R}^{m \times 2}$ is sparse, with nonzero entries only at the boundary nodes:
\begin{equation}\nonumber
    B = \frac{2 \Delta t}{h} \tilde{B}, \qquad \tilde{B}_{1,1} = 1, \quad \tilde{B}_{m,2} = 1,
\end{equation}
and all other entries of $\tilde{B}$ equal to zero. The stability condition for the forward Euler scheme requires $r \leq 1/2$. To generate data, we simulate $200$ trajectories using the above discretization. For each trajectory, we sample the initial condition from a Gaussian distribution $X_0\sim\mathcal{N}(0, I_m)$, apply i.i.d. uniform boundary controls $U_n\sim\operatorname{Unif}([-u_{\text{max}}, u_{\text{max}}]^2)$ with $u_{\text{max}}=5$ independent of state, and record pairs $(X_n, U_n, X_{n+1})$ over $200$ steps.

For dependence analysis, we model the distributions of $\Delta x_{1,n}$ and $\Delta x_{100, n}$ individually where $\Delta x_{i, n} := x_{i, n+1} - x_{i, n}$, and identify their sparse dependence structures from the full candidate input set $\{x_{1, n}, ..., x_{m, n}, u_{1, n}, u_{2, n}\}$. We use an FNN with four hidden layers of 64 neurons each and SiLU activation function to parameterize the velocity field, with batch size and number of epochs both set to $1{,}000$. We use $5{,}000$ distilled samples and the RBF kernel with lengthscales $1.0$ for $x_i, i=1,...,100$, $u_1, u_2$, and $z$.

As discussed in the main text, we evaluate control by replacing the boundary-node dynamics with data-driven surrogates inside model predictive control (MPC), while interior nodes keep the known finite-difference update. We use all training trajectories to train two conditional flow matching models using the recovered sparse input sets, yielding boundary maps $\Delta x_{1,n} = G_1(x_{1, n}, x_{2, n}, u_{1, n}, z_{1, n})$ and $\Delta x_{100,n} = G_{100}(x_{99, n}, x_{100, n}, u_{2, n}, z_{100, n})$ where $G_i$ denotes the time-1 flow generated by integrating the learned velocity field. 
Because repeated ODE integration is too costly for online MPC, we distill each teacher into an FNN by regressing the student output to the teacher map with MSE loss. Each student is a three-layer FNN (128 hidden units, SiLU activation function), trained for 20,000 steps with Adam ($1\times 10^{-3}$ learning rate) using mini-batches of 512 synthetic data pairs labeled by the teacher. 



\subsubsection{Details of the MPC setup}
Recall that in this example, the control objective is to regulate the temperature
profile to zero under a quadratic cost. For each independent trial, the initial
temperature profile is sampled as
\[
X_0 = 100 + 0.1\epsilon,
\qquad \epsilon \sim \mathcal{N}(0,I_m),
\]
and the same initial condition is used for all three controllers within that trial.
Over $10{,}000$ steps, $n=0,\ldots,9{,}999$, we solve a finite-horizon stochastic optimal control problem using a receding-horizon strategy. Specifically, given the current state $X_n$, we minimize a scenario-averaged finite-horizon quadratic cost over $U_{n:n+H-1}$:
\begin{equation}
    \min_{U_n,\ldots,U_{n+H-1}}
    \frac{1}{S}\sum_{s=1}^{S}
    \left[
    \sum_{k=0}^{H-1}
    \left(
    \left\|X_{n+k}^{(s)}\right\|_Q^2
    +
    \left\|U_{n+k}\right\|_R^2
    \right)
    +
    \left\|X_{n+H}^{(s)}\right\|_Q^2
    \right],
    \label{eq:mpc_objective}
\end{equation}
where $s=1,\ldots,S$ indexes the scenarios, and $\|v\|_M^2:=v^\top Mv$.
Here, $Q=I_m\in\mathbb{R}^{m\times m}$ penalizes deviation of the temperature field from zero, and $R=\rho I_2\in\mathbb{R}^{2\times2}$ penalizes control effort, with $\rho=0.01$. Each $U_{n+k}$ is clipped to $[-5,5]^2$ at each step. The expectation over the stochastic dynamics is approximated by averaging over $S$ independent scenarios. Only the first control $U_n^*$ is applied to the true system, after which the optimization is re-solved at $n+1$.

We assume that inside the MPC rollout, the update functions for the two boundary nodes, $\Delta x_1$ and $\Delta x_{100}$, are unknown and must be learned from data. The dynamics of the remaining interior nodes $\Delta x_i$, $i=2,\ldots,m-1$, are assumed known exactly. The boundary updates are represented by stochastic generative models.
We compare three controllers, all targeting the same objective~\eqref{eq:mpc_objective}:
\begin{enumerate}
    \item \textbf{MPC with known dynamics}: the exact boundary dynamics are used as a reference.
    \item \textbf{MPC with full surrogate}: the dynamics of $\Delta x_1$ and $\Delta x_{100}$ are modeled conditioned on the full input set $\{x_{1,n},\ldots,x_{m,n},u_{1,n},u_{2,n}\}$.
    \item \textbf{MPC with sparse surrogate}: the dynamics of $\Delta x_1$ and $\Delta x_{100}$ are modeled conditioned on the recovered sparse input sets, $\{x_{1, n},x_{2, n},u_{1, n}\}$ for $\Delta x_1$ and $\{x_{99, n},x_{100, n},u_{2, n}\}$ for $\Delta x_{100}$.
\end{enumerate}
The three controllers use the same MPC objective, optimization procedure, and interior dynamics, and differ only in the boundary dynamics used during the rollout. The optimization problem~\eqref{eq:mpc_objective} is solved using the L-BFGS method~\cite{liu1989limited} with the maximum number of iterations set to $300$. The control sequence is warm-started at each time step by shifting the previous solution.
We solve the MPC optimization over a planning horizon of $H=20$ steps with $S=64$ Monte Carlo scenarios, and evaluate over ten independent trials.

\subsection{The economic data example}\label{appendix:economic}

Before constructing the dataset, we synchronize the oil-price and geopolitical risk (GPR) index time series by retaining only dates for which both quantities are available. Since oil prices are reported only on trading days while the GPR index is available daily, all non-trading days are removed from the GPR series. Dates containing missing values or extreme anomalous observations (e.g., negative prices) in either variable are also discarded. The resulting dataset therefore consists of a filtered sequence of aligned observation dates. The oil-price return is then defined on this filtered sequence as
\begin{equation}
    r_n = \log(\text{Oil}_{n+1}) - \log(\text{Oil}_{n}),
\end{equation}
where $\text{Oil}_n$ denotes the oil price at the $n$th retained observation date in the filtered dataset. Consequently, the return is computed between consecutive retained observations rather than strictly between consecutive calendar days.

We note that the selected lag should not be interpreted as the unique physical response time of the oil market to geopolitical risk. The lag candidates are user-defined coarse representatives of delayed information channels. Changes in the dominant surviving lag instead indicate structural changes in the dependence structure between geopolitical risk and oil prices across different market regimes.

In addition to the three sub-periods analyzed in the main text, we consider eight shorter sub-periods (the start and end years are inclusive): 1987--1991, 1992--1996, 1997--2001, 2002--2006, 2007--2011, 2012--2016, 2017--2021, 2022--2026/04/20. We apply the proposed method to analyze the structure of the transition kernel. For each period, we focus on the GPR lag with the largest surviving contribution score in modeling the transition kernel in \eqref{eq:economic}, defined as the variable removed last by the proposed method.

We find that the 1-day lag is the dominant surviving variable in 1987--1991, 1992--1996, 2002--2006, 2017--2021, and 2022--2026/04/20; the 10-day lag in 1997--2001 and 2007--2011; and the 5-day lag in 2012--2016. In particular, the dominant lag changes from 1-day in 1987--1996 to 10-day in 1997--2001, back to 1-day in 2002--2006, and again to 10-day in 2007--2011. This pattern is consistent with the longer-period analysis presented in the main text.
The periods 1997--2001 and 2007--2011 overlap with major geopolitical and economic disruptions, including the 9/11 attacks and the 2008 financial crisis. We do not attribute the detected changes in dependence structure to these individual events, but note that the identified shifts occur during periods of substantial geopolitical and economic disruption.

We also observe that the 2012--2016 period, for which the 5-day lag is dominant, overlaps with substantial changes in oil-market conditions, including shale supply expansion and changes in OPEC policy. These temporal coincidences provide context for the detected changes but should not be interpreted as establishing causal relationships.

Overall, these results suggest that the delayed dependence structure between geopolitical risk and oil returns is not stationary across time, and that the proposed method can be used to identify and localize changes in stochastic dependence structure directly from observational economic data.

In all cases, we use an FNN with four hidden layers of $64$ neurons each and SiLU activation function to parameterize the velocity field, trained with batch size $1{,}000$ for $5{,}000$ epochs. We then use $10{,}000$ distilled samples and an RBF kernel with lengthscale $5.0$ for both the state and latent variables. The GP nugget is set to $\exp(1.0)$. Only the GPR lags are considered for removal.

\section{Additional examples}\label{sec:appendix:additional_examples}

In this section, we provide additional examples that further validate the proposed dependence-discovery framework across a range of stochastic systems.

\subsection{Illustrative example: dependence discovery and sparse stochastic prediction}

\begin{figure}[t]
\centering\includegraphics[width=1.0\linewidth]{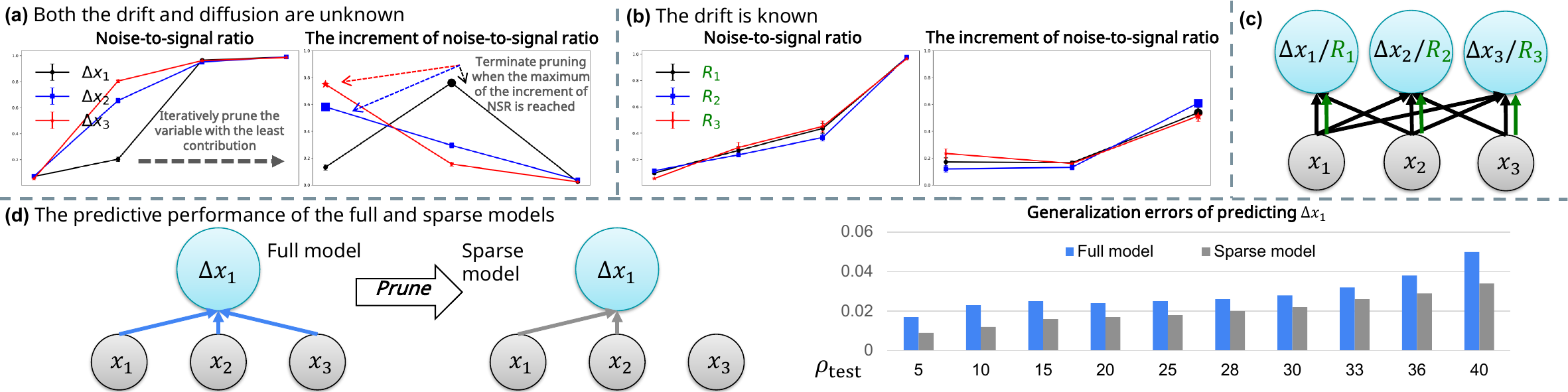}
    \caption{Stochastic Lorenz-63 system. (a)--(b): pruning process over ten independent runs. (c): Recovered dependence structure. (d): Generalization of the full and sparse models under parameter-induced distribution shift by training at $\rho=28$ and testing at different values of $\rho_{\mathrm{test}}$.
    }
    \label{fig:lorenz63}
\end{figure}

We first use the stochastic Lorenz-63 system as an illustrative example to demonstrate how the proposed dependence-discovery and variable-removal procedure operates in practice. The system is governed by multiplicative fluctuation-dissipation noise \cite{lorenz1963deterministic, geurts2020lyapunov}:
\begin{subequations}
\begin{align}
dx_{1} &= \sigma(x_{2} - x_{1})dt + \alpha x_{1}dW_{1},\\
dx_{2} &= \bigl(x_{1}(\rho - x_{3}) - x_{2}\bigr)dt + \alpha x_{2}dW_{2},\\
dx_{3} &= (x_{1}x_{2} - \beta x_{3})dt + \alpha x_{3}dW_{3},
\end{align}
\end{subequations}
where $x_i$ are the state variables and $W_i$ are independent Wiener processes. We recover the dependence structure of the state increments by applying the proposed method to samples of $\Delta x_i, i=1,2,3$. We consider two cases. First, both drift and diffusion are treated as unknown. In this case, the targets are the full increments $\Delta x_i$. 
The reference dependence sets are $\{x_1,x_2\}$ for $\Delta x_1$, and $\{x_1,x_2, x_3\}$ for $\Delta x_2$ and $\Delta x_3$, obtained by inspecting both the drift and diffusion terms. Second, when the drift is known, we apply the method to the residual stochastic increments $R_i := \Delta x_i - b_i(x_1,x_2,x_3)\Delta t$, where $b_i$ is the $i$th drift component. Since the diffusion is diagonal, the reference dependence sets are $\{x_1\}, \{x_2\}$, and $\{x_3\}$. 

Figures~\ref{fig:lorenz63}(a)--(b) illustrate the iterative variable-removal procedure over ten independent runs. At each step, the variable with the smallest contribution score is removed and the reduced model is redistilled using the remaining inputs. The NSR remains nearly unchanged while variables with negligible contribution are removed, indicating that the reduced input set still provides an accurate representation of the learned conditional law. Once removal eliminates a variable required to represent the conditional law, the NSR increases sharply. We therefore terminate the procedure immediately before the largest NSR increase. Figure~\ref{fig:lorenz63}(c) shows that the resulting sparse input sets agree with the reference dependence structure in both settings.

Beyond validating the variable-removal procedure, this example also illustrates why removing unnecessary inputs can improve stochastic surrogate modeling. We focus on the conditional law of $\Delta x_1$, which does not depend on $x_3$ according to the governing equation. We compare a full model conditioned on $(x_1,x_2,x_3)$ with a sparse model conditioned only on $(x_1,x_2)$. Both models are trained using data generated at $\rho_{\mathrm{train}}=28$ and then evaluated for one-step prediction using data generated at a range of values of $\rho_{\mathrm{test}}$. As shown in Figure~\ref{fig:lorenz63}(d), the sparse model consistently achieves lower prediction error, measured by the Wasserstein-1 distance, than the full model, both in the training regime and under parameter-induced distribution shift. Thus, even in this low-dimensional setting, removing a single input on which the conditional law does not depend can improve the accuracy and robustness of the learned stochastic model.

\subsubsection{Additional details}
In this example, we use the standard parameters $\sigma=10$, $\rho=28$, $\beta=8/3$, noise amplitude $\alpha=0.5$, and time step $\Delta t=0.01$. Initial conditions are sampled independently from $\mathrm{Unif}([-15,15]^3)$. The system is simulated with the Euler--Maruyama method. We generate $50$ independent trajectories; each trajectory is run for $200$ burn-in steps and then contributes $1{,}000$ state-increment pairs.

To assess run-to-run robustness, we repeat the experiment ten times independently. In both settings, all ten runs recover the same graphs. The NSR and NSR-increment curves in Figure~\ref{fig:lorenz63}(a)--(b) report the mean over the ten runs, with error bars indicating one standard deviation.
The target-wise pruning orders are also largely consistent: for $\Delta x_2$, $R_2$, and $R_3$, all ten runs yield the same pruning orders, while for $\Delta x_1$, $\Delta x_3$, and $R_1$, nine, nine, and eight runs yield the same pruning orders, respectively.  

We also use this example to assess the approximation error introduced by the two-stage architecture: a flow matching model is first learned from data, then distilled into a GP surrogate on which KMD is performed. Specifically, we evaluate the GP surrogate in two ways: first, by its regression accuracy on distilled samples $\{(X^i,z^i,\hat y^i)\}_{i=1}^M$ generated from the learned conditional sampler; and second, by the accuracy with which it reproduces the corresponding conditional distributions on test inputs generated from the same data-generating process.
Results for $\Delta x_{1, n}$ are presented in Table~\ref{tab:lorenz63}. As $M$ increases, the Wasserstein-1 distance of the GP surrogate approaches that of the trained flow matching model, indicating that the GP distillation stage introduces little additional error when sufficiently many distilled samples are used. The number of validation distilled samples is $10{,}000$. 
We further examine the effect of the number of distilled samples $M$ on dependence discovery. The GP surrogate is trained separately for each target variable $\Delta x_{k,n}$. For $\Delta x_{1,n}$ and $\Delta x_{3,n}$, the correct dependence structures are recovered for all values of $M$ tested. For $\Delta x_{2,n}$, $M=500$ and $M=2{,}000$ yield the reduced set $\{x_{1,n}, x_{3,n}\}$, missing $x_{2,n}$, while $M=5{,}000$ and $M=10{,}000$ recover the complete dependence set $\{x_{1,n}, x_{2,n}, x_{3,n}\}$. This reflects the fact that the dependence of $\Delta x_{2,n}$ on $x_{2,n}$ is the weakest among the three components, and a sufficiently faithful GP surrogate is required to resolve it.

\begin{table}[h]
    \footnotesize
    \centering
    \begin{tabular}{c|c|c|c|c}
    \hline
    \hline
     & $M=500$ & $M=2{,}000$ & $M=5{,}000$ & $M=10{,}000$\\
    \hline
    MSE on validation distilled samples & 6.02e-2 & 2.08e-2 & 8.34e-3 & 5.19e-3 \\
    \hline
    The Wasserstein-1 distance (mean$\pm$std) & 7.47e-2$\pm$1.23e-1 & 4.09e-2$\pm$6.54e-2 & 3.31e-2$\pm$3.82e-2 & 3.11e-2$\pm$3.30e-2 \\
    \hline
    \hline
    \end{tabular}
    \caption{Approximation quality of the GP surrogate for the generative model of $\Delta x_{1, n}:=x_{1, n+1} - x_{1, n}$. \textit{Top row}: the regression mean-squared error (MSE) on validation distilled samples generated from the learned conditional sampler. \textit{Bottom row}: Wasserstein-1 distance to the exact data-generating process, evaluated on $1{,}000$ test inputs with $1{,}000$ surrogate samples drawn per test point. For reference, the Wasserstein-1 distance of the trained flow matching model to the exact data-generating process is 3.42e-2$\pm$2.55e-2.}
    \label{tab:lorenz63}
\end{table}

Here, we model the conditional distributions of the increments $\Delta x_{k,n}$,
$k=1,2,3$, given the current state $(x_{1,n},x_{2,n},x_{3,n})$.
We use an FNN with four hidden layers of $64$ neurons each and sigmoid linear unit (SiLU) activation function to parameterize the velocity field of the flow matching model, with batch size and number of epochs both set to $1{,}000$.
In the second stage, we use $M=5{,}000$ distilled samples $\{x^i_{k,n}, z^i_k, \hat{\Delta x_{k, n}}, k=1,2,3\}_{i=1}^{5{,}000}$ to train the GP surrogate. We use the RBF kernel with lengthscale $1.0$ for both the state and the latent variables.

\subsection{Recovering the dependence structure of a stochastic differential equation}

This example serves as a verification test for the proposed method since both the conditional distributions and the underlying dependence structure are known. Specifically, we consider the stochastic Kraichnan–Orszag (KO) system \cite{wan2006multi}:
\begin{subequations}
\begin{align}
dx_1 &= x_1x_3\,dt + \sigma_1\,dW_1,\\
dx_2 &= -x_2x_3\,dt + \sigma_2\,dW_2,\\
dx_3 &= (-x_1^2 + x_2^2)\,dt + \sigma_3\,dW_3,
\end{align}
\end{subequations}
where $W_i$ are independent Wiener processes. The system is simulated using the Euler--Maruyama method with randomly sampled initial conditions and noise parameters $\sigma_1=\sigma_2=\sigma_3=0.1$.

For dependence discovery, we consider the discretized update equations
\begin{subequations}
\begin{align}
x_{1,n+1} &= x_{1,n} + x_{1,n}x_{3,n}\Delta t + \sigma_1\sqrt{\Delta t}\xi_1,\\
x_{2,n+1} &= x_{2,n} - x_{2,n}x_{3,n}\Delta t + \sigma_2\sqrt{\Delta t}\xi_2,\\
x_{3,n+1} &= x_{3,n} + (-x_{1,n}^2 + x_{2,n}^2)\Delta t + \sigma_3\sqrt{\Delta t}\xi_3,
\end{align}
\end{subequations}
and define $\Delta x_{i,n} := x_{i,n+1}-x_{i,n}$. The goal is to learn stochastic update functions
\[
\Delta x_{i,n} \approx \Phi_i(x_{1,n},x_{2,n},x_{3,n},\omega_i), \qquad i=1,2,3.
\]

\begin{figure}
    \centering
    \subfigure[]{
        \includegraphics[width=0.5\linewidth]{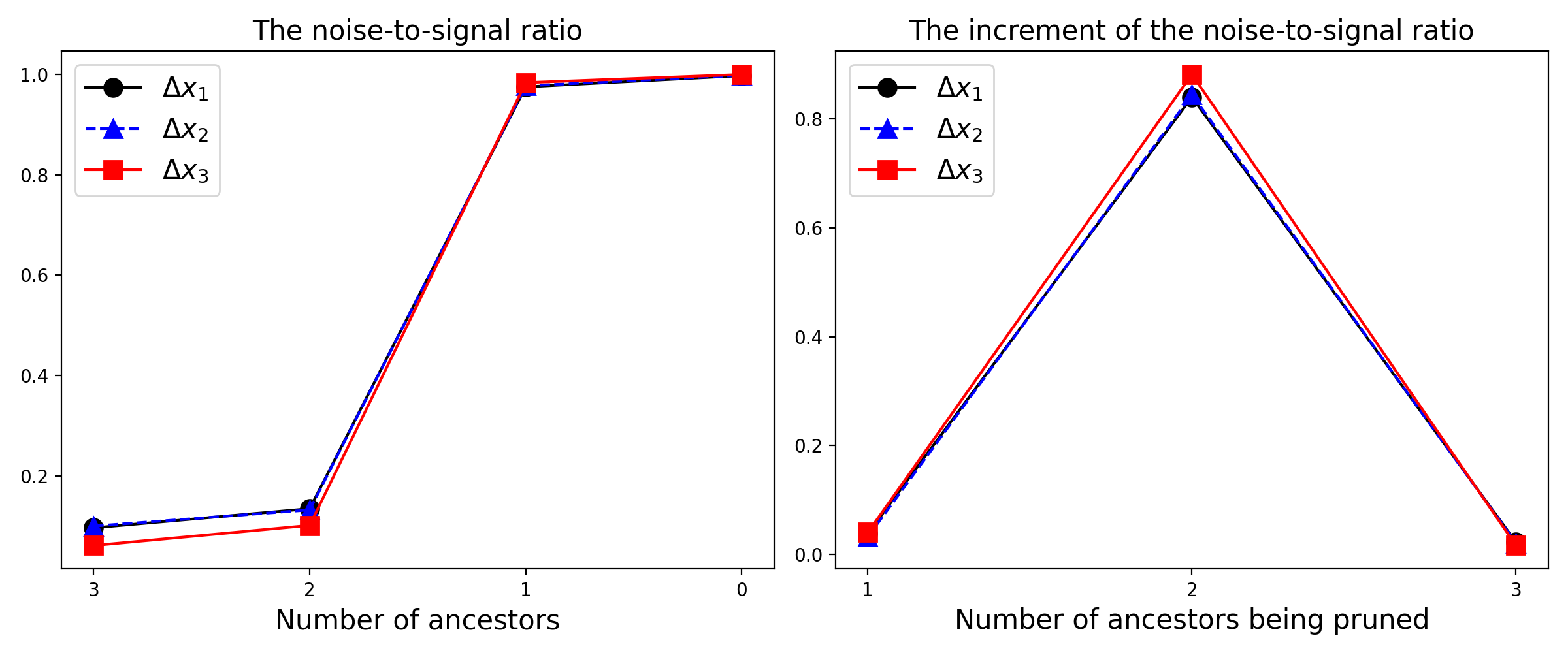}
    }
    \subfigure[]{
        \includegraphics[width=0.4\linewidth]{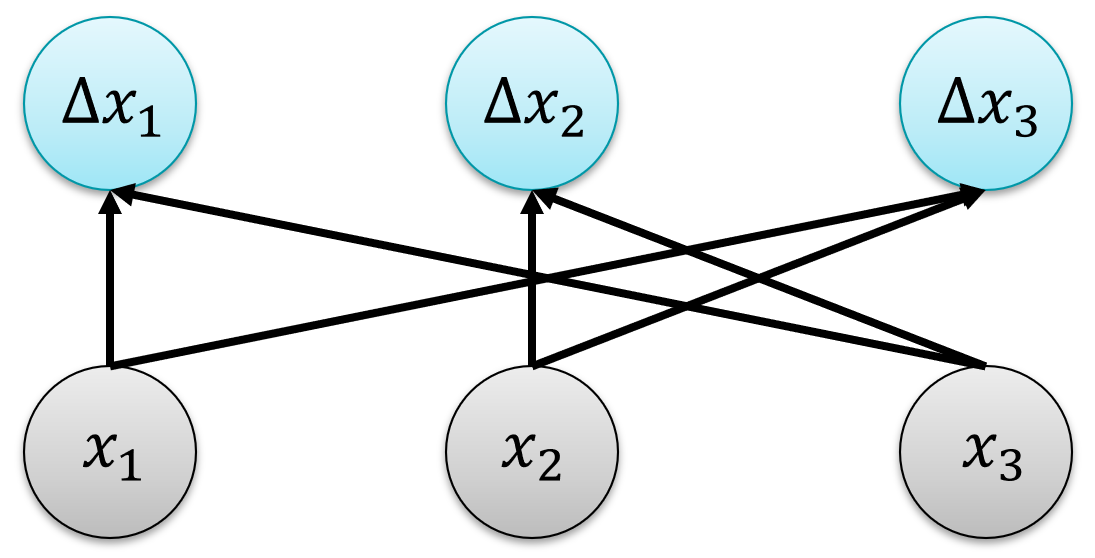}
    }
    \caption{Stochastic KO system. (a): Iterative variable-removal process. (b): Recovered dependence structure.}
    \label{fig:ko_1}
\end{figure}

Figure~\ref{fig:ko_1} shows the iterative variable-removal process and the recovered dependence structure. To evaluate the accuracy of the learned generative models, we compute the Wasserstein-1 distance between the learned conditional distributions and the exact distributions. The results are reported in Table~\ref{tab:example_1}. The comparison shows that using the sparse input set yields accuracy comparable to using all candidate variables, indicating that the proposed method successfully identifies the correct dependence structure without degrading generative modeling performance.

\begin{table}[h]
    \footnotesize
    \centering
    \begin{tabular}{c|c|c|c}
    \hline
    \hline
     & $\Phi_1$ & $\Phi_2$ & $\Phi_3$\\
    \hline
    Using all input variables (1e-4) & $1.15\pm1.30$ & $1.16\pm0.60$ & $1.59\pm2.06$ \\
    \hline
    Using sparse input variables (1e-4) & $0.95\pm0.82$ & $0.92\pm0.46$ & $1.40\pm1.84$ \\
    \hline
    \hline
    \end{tabular}
    \caption{Average Wasserstein-1 distance between the learned generative update model and the exact conditional distribution. We randomly select $1{,}000$ test data points. For each one of them, $W_1$-distance is estimated based on $10{,}000$ samples.} 
    \label{tab:example_1}
\end{table}

To generate data, we use $\Delta t=0.001$. Initial conditions are sampled from $\mathcal{N}(0,2^2I_3)$, where $I_3$ denotes the $3\times3$ identity matrix, and the system is simulated with the Euler--Maruyama method. We generate $50$ independent trajectories; each trajectory is run for $10{,}000$ steps, yielding $500{,}000$ state-increment pairs in total. We model the conditional distributions of the increments $\Delta x_{k,n}$, $k=1,2,3$, separately, using the current state $(x_{1,n},x_{2,n},x_{3,n})$ as candidate inputs. We use an FNN with two hidden layers of $64$ neurons each and SiLU activation function to parameterize the velocity field of the flow matching model, with batch size $5{,}000$ and number of epochs $500$. In the second stage, we use $M=5{,}000$ distilled samples to fit the GP surrogate. We use the RBF kernel with lengthscale $1.0$ for both the state and latent variables.

\subsection{Discovering memory dependence in stochastic dynamics}

We consider a coarse-grained stochastic dynamical system in which the momentum is hidden, so the observed position dynamics exhibit memory. The underlying system follows a Langevin equation for a double-well potential
\begin{subequations}
    \begin{align}
        \frac{dx}{dt} &=p, \\
        \frac{dp}{dt} &=-V'(x)-\alpha p+\sigma\xi(t),
    \end{align}
\end{subequations}
where
$$
V(x)=V_0\left(1-\left(\frac{x}{x_0}\right)^2\right)^2
$$
is a symmetric double-well potential with two metastable states \cite{zhu2023learning, liu2015equation}.

\begin{figure}
    \centering
    \subfigure[]{
        \includegraphics[width=0.5\linewidth]{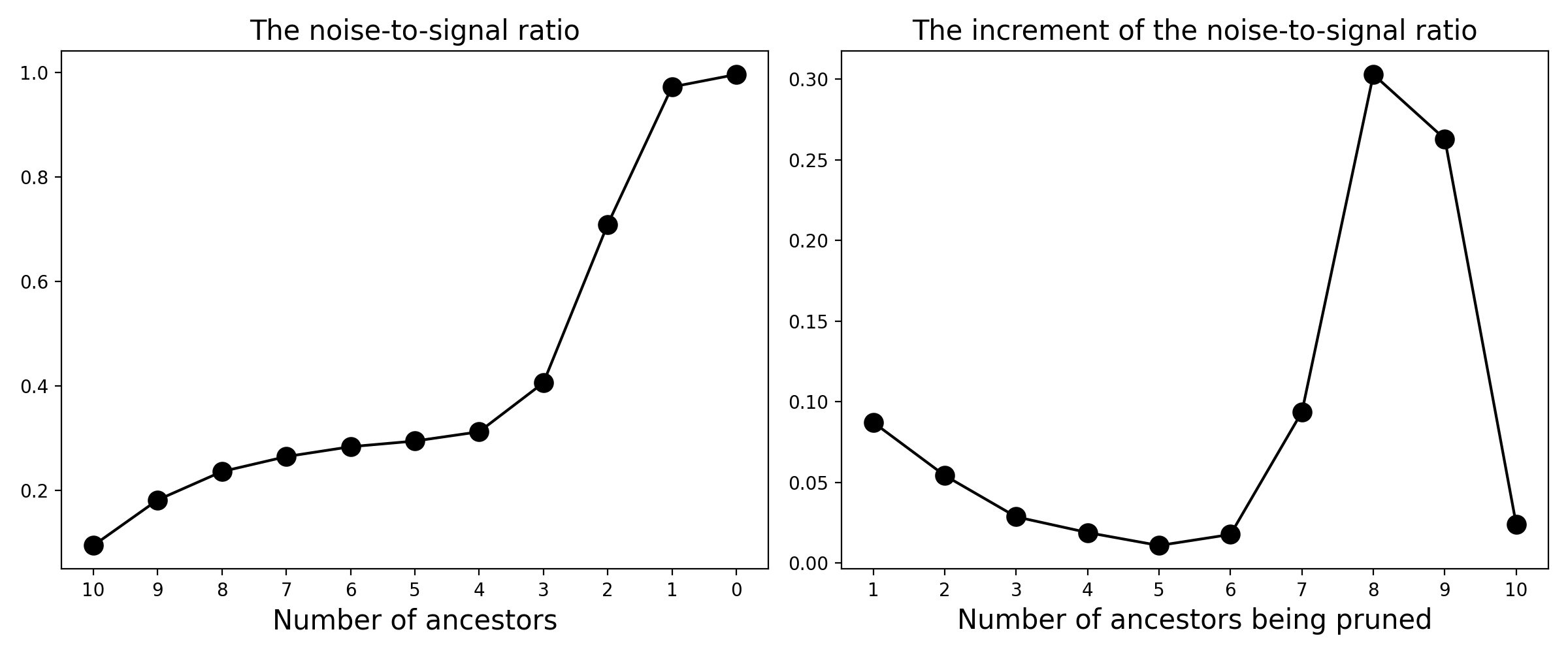}
    }
    \subfigure[]{
        \includegraphics[width=0.25\linewidth]{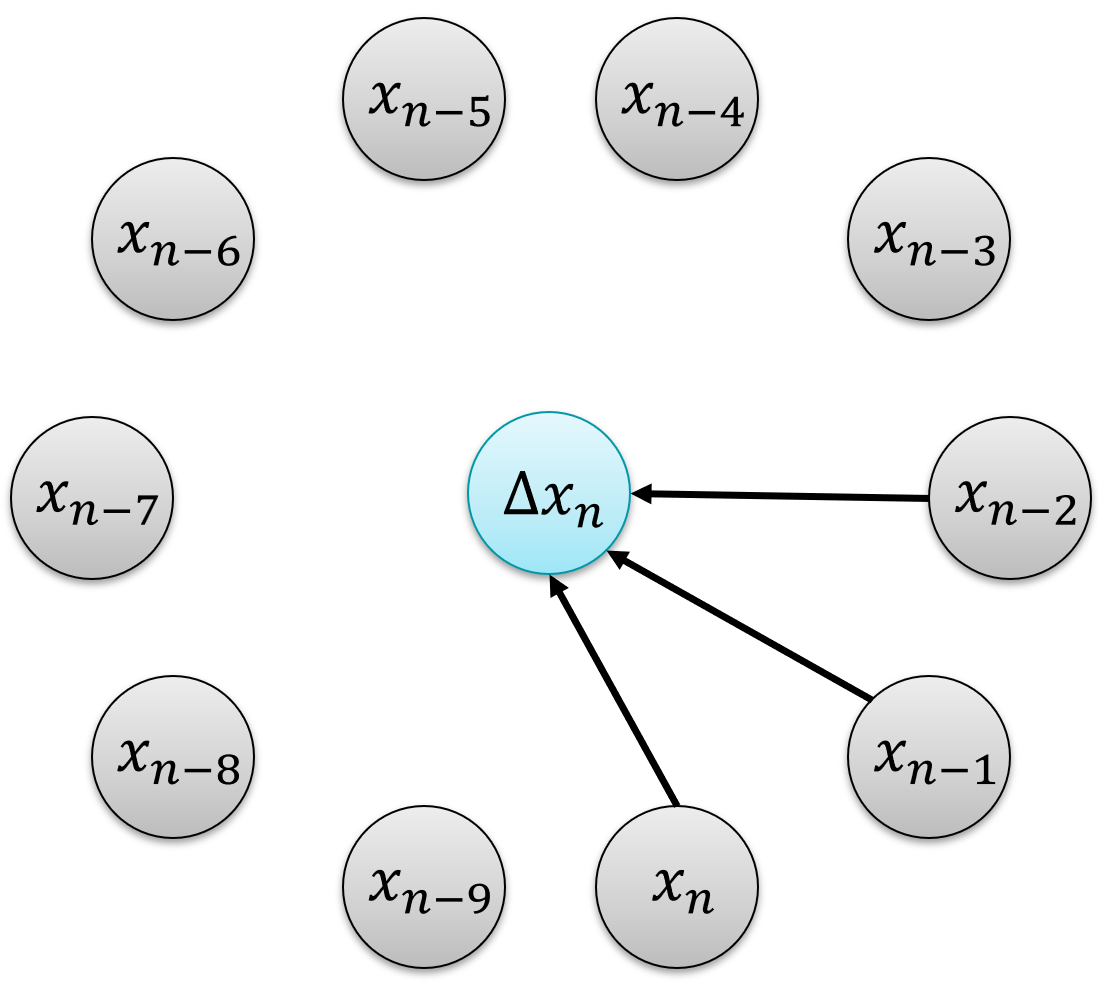}
    }
    \subfigure[]{
        \includegraphics[width=0.2\linewidth]{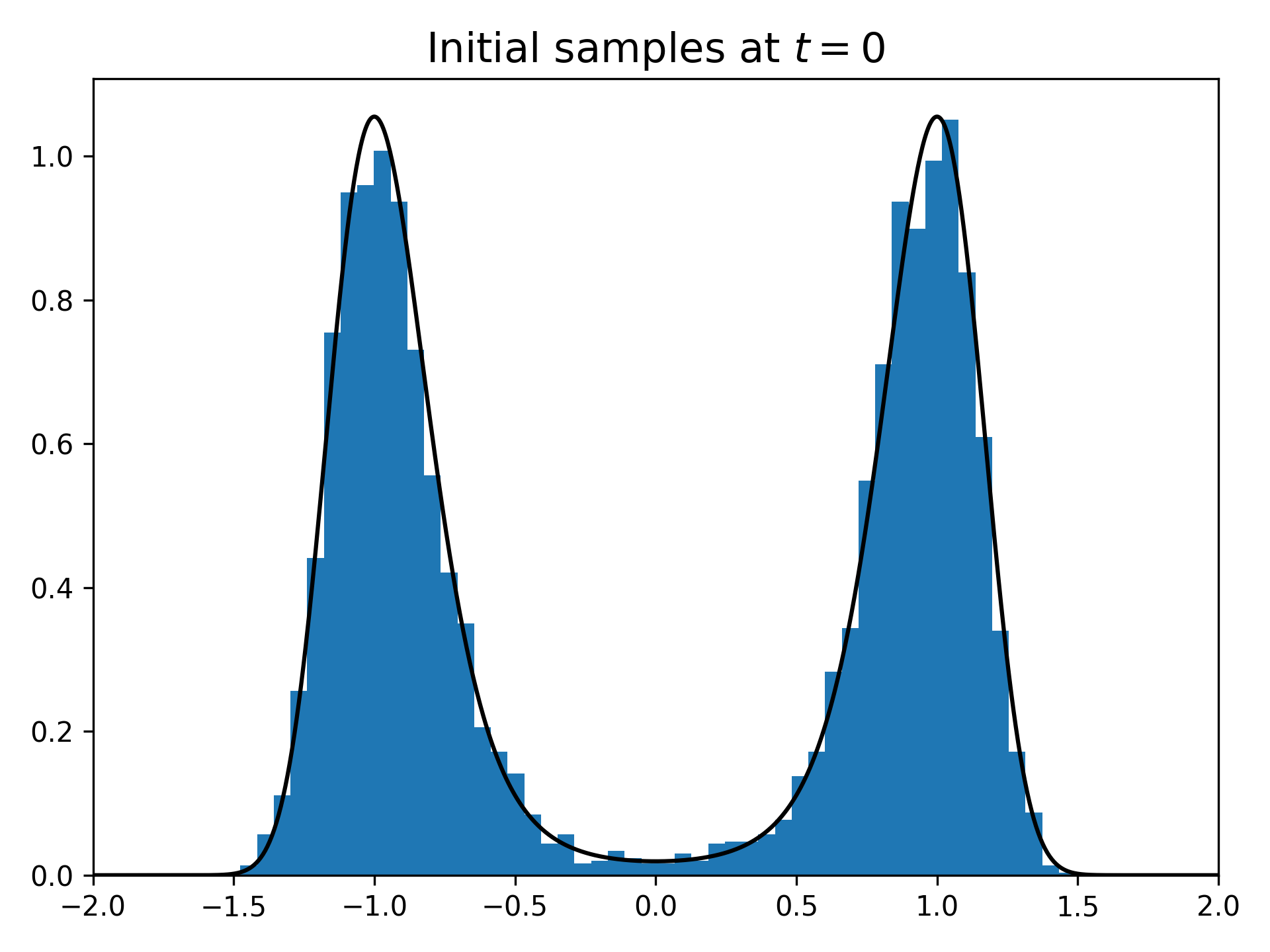}
        \includegraphics[width=0.2\linewidth]{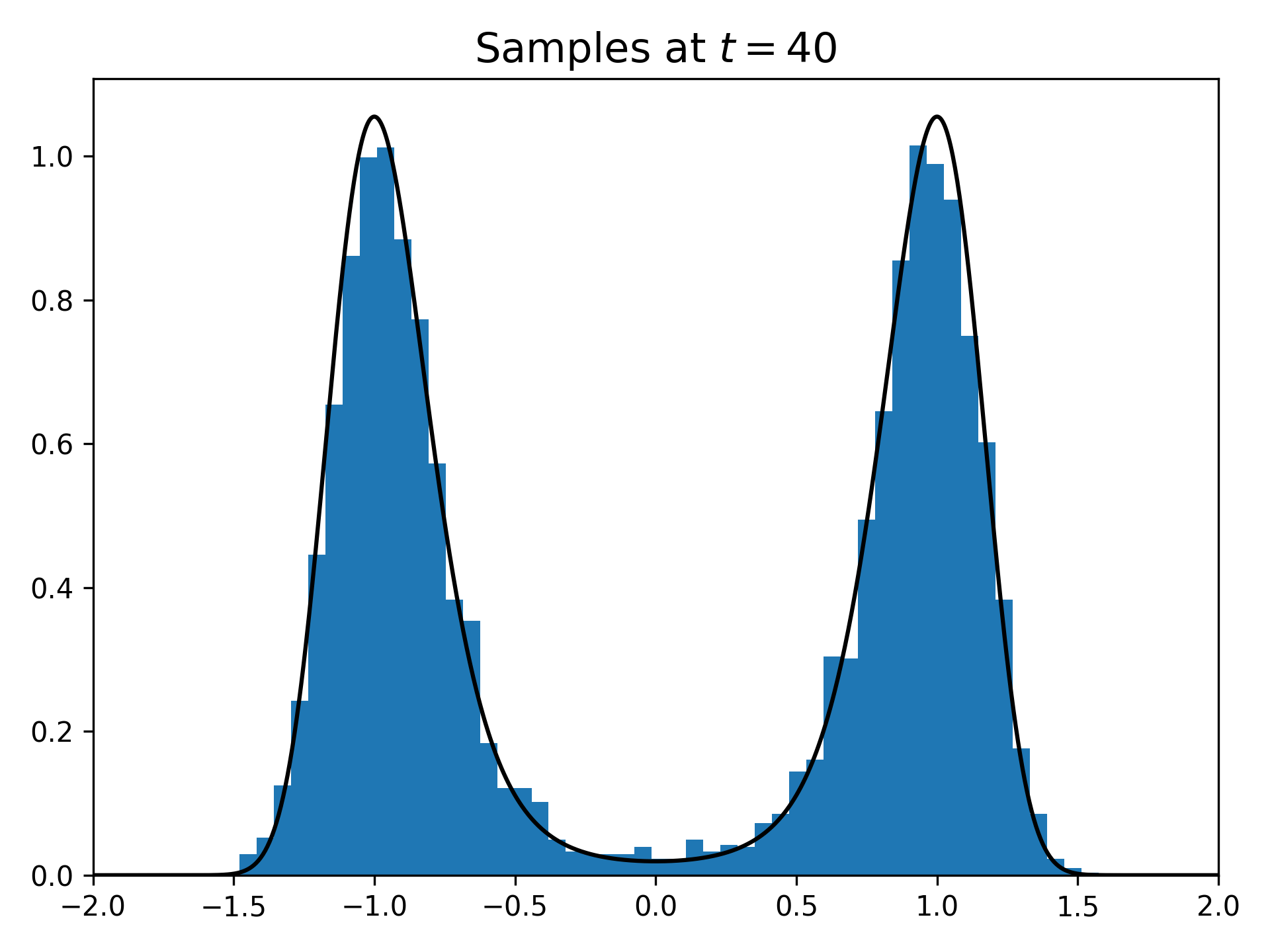}
        \includegraphics[width=0.2\linewidth]{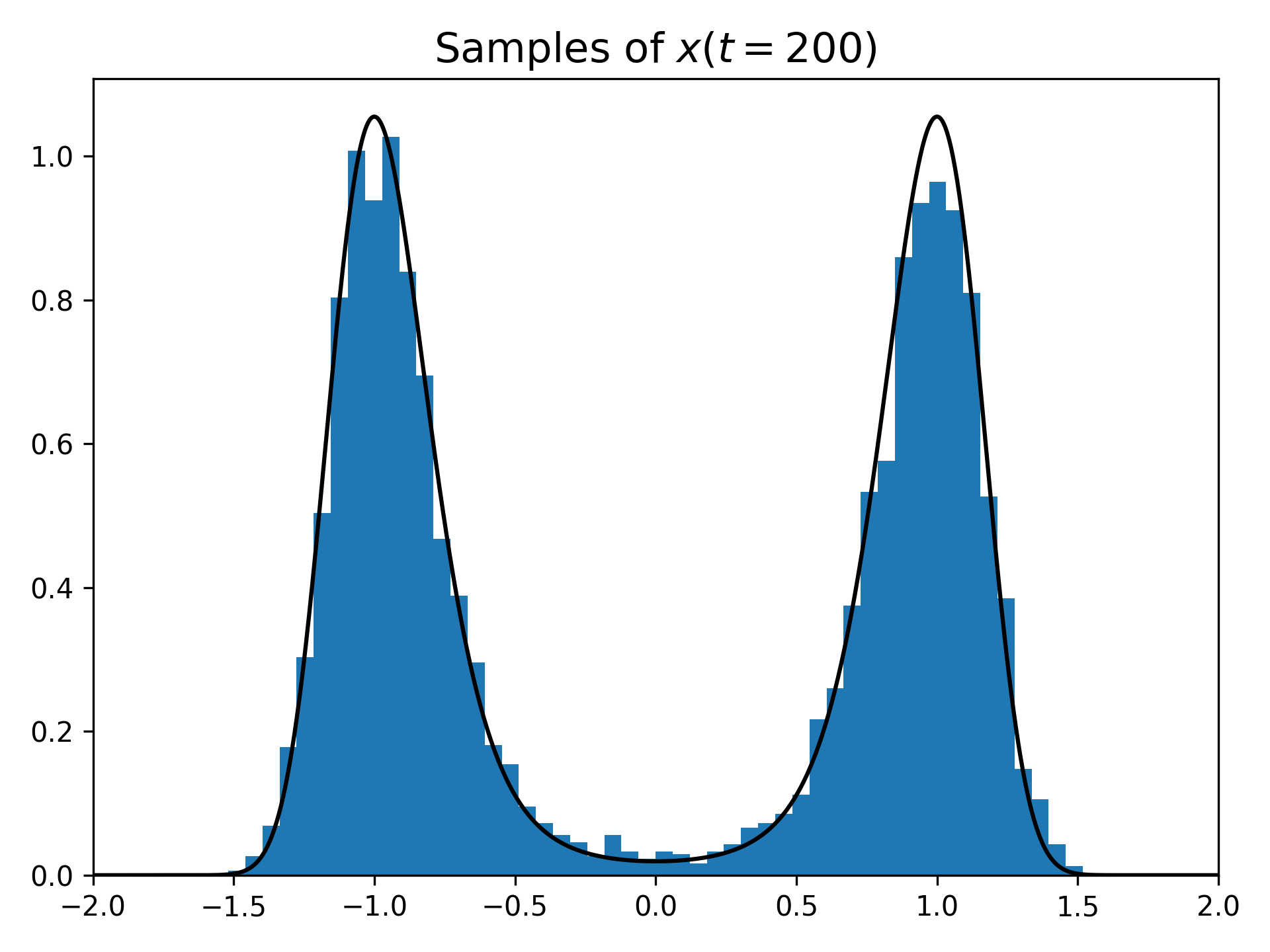}
    }
    \subfigure[]{
        \includegraphics[width=0.2\linewidth]{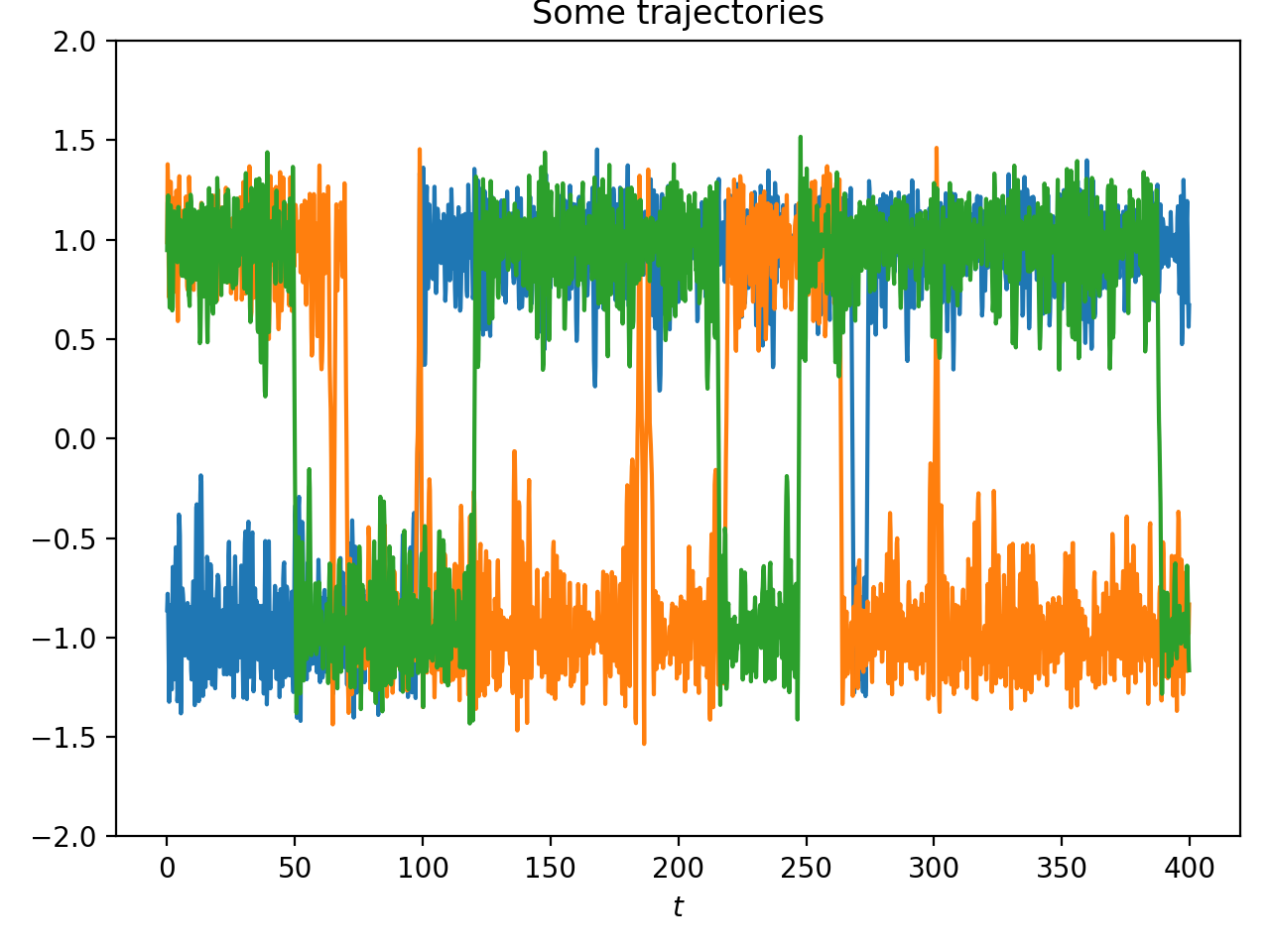}
    }
    \caption{Coarse-grained transition dynamics with a missing variable. (a): Iterative variable-removal process. (b): Discovered dependence structure. (c): Generated samples of $x$ at different times from the learned conditional law. (d): Generated trajectories from the learned conditional law.}
    \label{fig:example_2_1}
\end{figure}

In this example, we assume that only the position variable $x$ is observed while the momentum variable $p$ is hidden. The resulting coarse-grained dynamics of $x$ exhibit memory effects. We therefore model the dynamics using a stochastic update rule
\begin{equation}
x_{n+1}=x_n+\Phi(x_n,x_{n-1},\dots,x_{n-m},\omega_n),
\end{equation}
where the goal is to identify which lagged positions are needed to represent the transition law. Applying the proposed method allows us to determine the relevant memory states. Figure~\ref{fig:example_2_1} shows the iterative variable-removal process, the discovered dependence structure, and trajectories generated using the learned update model. 
The results show that the method identifies a compact set of lagged positions sufficient to reproduce the observed coarse-grained transition law.

We follow the setting in \cite{zhu2023learning} and set $\alpha=1$, $\sigma=\sqrt{2}$, $V_0=4$, and $x_0=1$. The system is simulated using the Euler--Maruyama method with $\Delta t=0.001$. We generate $10{,}000$ trajectories from the zero initial condition, each for $320{,}000$ time steps, and record the state every $200$ steps. The first $160{,}000$ steps are discarded as burn-in. We use $5{,}000$ trajectories for training and the remaining $5{,}000$ for testing.
For each training trajectory, we use the first $101$ saved positions after burn-in to construct the training data. With $m=9$, each sample contains the ten lagged positions $(x_n,x_{n-1},\ldots,x_{n-9})$, yielding $91$ samples per trajectory. We parameterize the velocity field of the flow matching model using an FNN with two hidden layers of $64$ neurons each and hyperbolic tangent activation, trained with batch size $5{,}000$ for $1{,}000$ epochs. In the second stage, we fit the GP surrogate using $M=5{,}000$ distilled samples and an RBF kernel with lengthscales $1.0$ and $10.0$ for the state and latent variables, respectively.

\subsection{Discovering state dependence in a stochastic chemical system}

\begin{figure}
    \centering
    \subfigure[]{
        \includegraphics[width=0.5\linewidth]{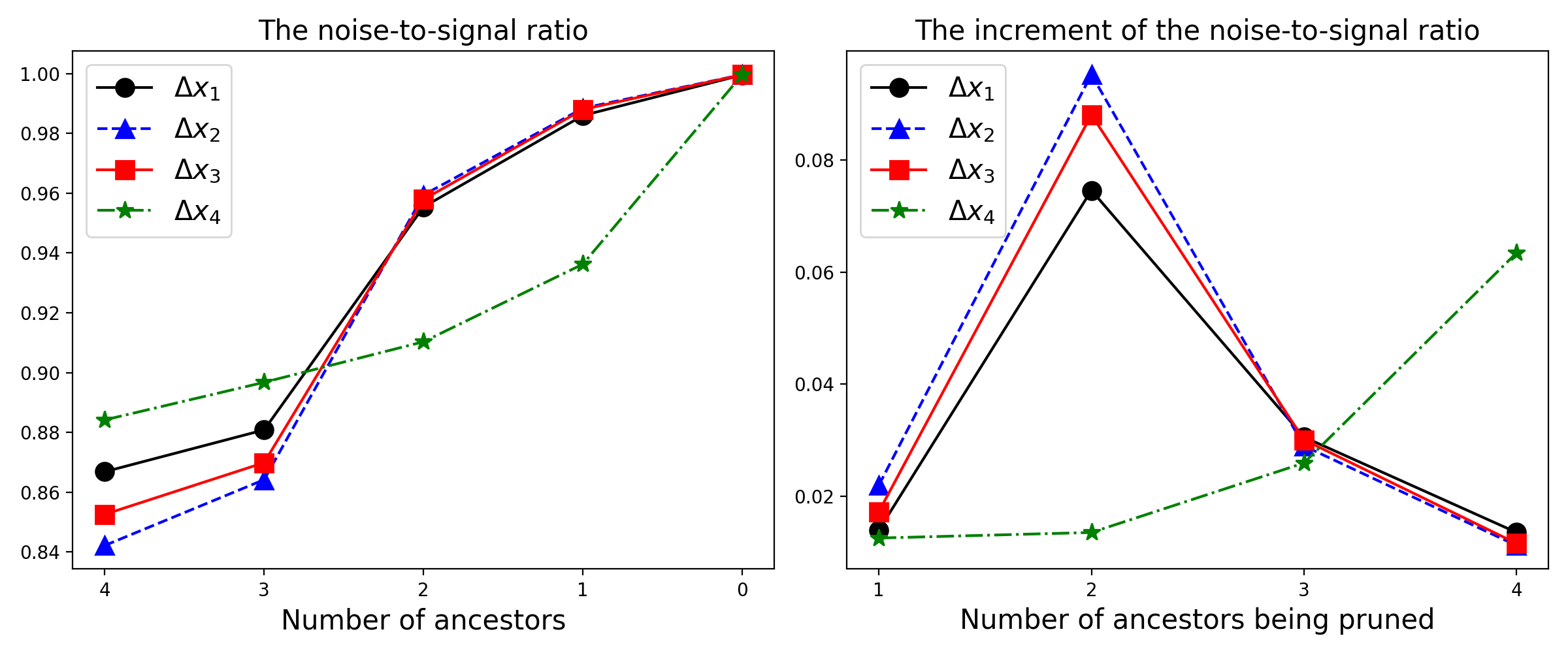}
    }
    \subfigure[]{
        \includegraphics[width=0.45\linewidth]{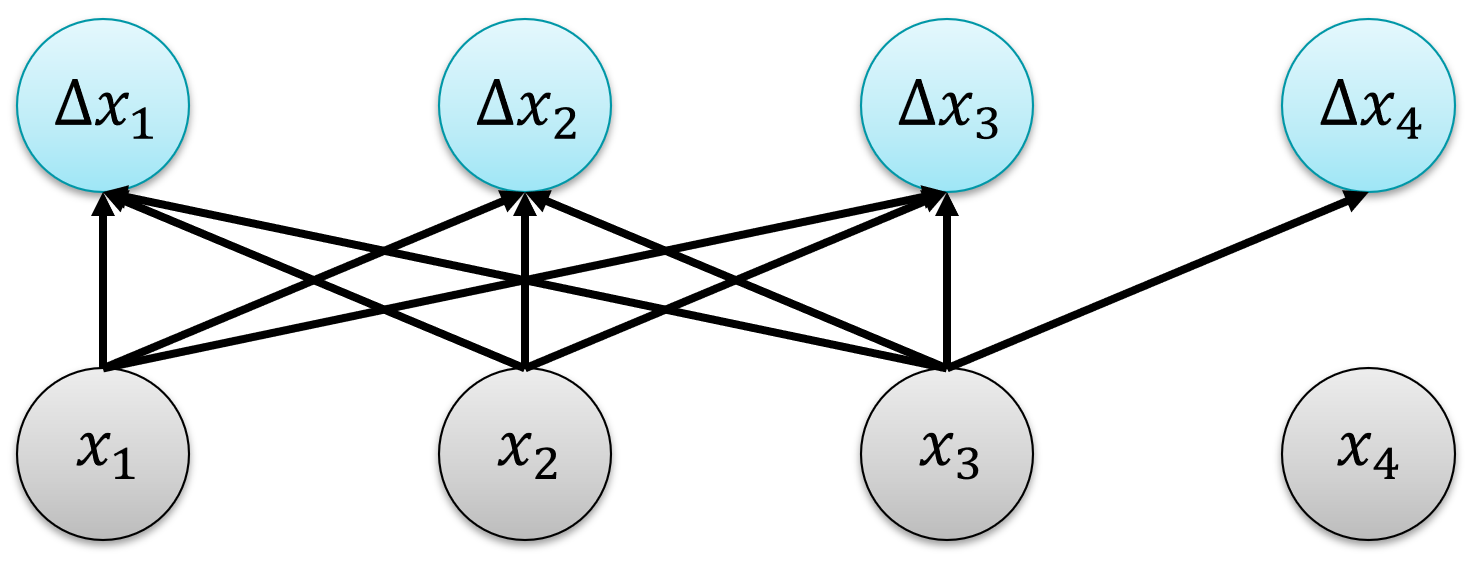}
    }
    \caption{Stochastic chemical system. (a): Iterative variable-removal process. (b): Discovered dependence structure.}
    \label{fig:chemical}
\end{figure}

This experiment illustrates the ability of the proposed method to discover state dependence in stochastic reaction networks from data generated by exact stochastic simulations. We consider the classical Michaelis--Menten enzymatic reaction mechanism involving four chemical species \cite{higham2008modeling}:
\[
S_1: \text{substrate}, \qquad
S_2: \text{enzyme}, \qquad
S_3: \text{complex}, \qquad
S_4: \text{product}.
\]
The reactions are
\begin{subequations}
\begin{align}
S_1 + S_2 &\xrightarrow{c_1} S_3,\\
S_3 &\xrightarrow{c_2} S_1 + S_2,\\
S_3 &\xrightarrow{c_3} S_4 + S_2,
\end{align}
\end{subequations}
where $c_1, c_2, c_3$ are the rate constants.
Let $X(t) = [x_1(t), x_2(t), x_3(t), x_4(t)]$ denote the state vector where $x_i(t)$ represents the number of molecules of species $S_i$ at time $t$. The stochastic reaction dynamics are simulated using Gillespie’s stochastic simulation algorithm (SSA) \cite{higham2008modeling}. At each step, a reaction is selected with probability proportional to its propensity function and the state is updated according to the corresponding stoichiometric change.

While SSA provides an exact simulation of the stochastic chemical kinetics, it does not directly yield a fixed-time update rule. To analyze the dependence structure using the proposed framework, we consider the discrete-time increments
\[
\Delta x_{i,n} = x_{i,n+1} - x_{i,n}, \qquad i=1,\dots,4,
\]
with a fixed observation interval $\tau=0.1$. The dataset consists of pairs $(X_n,\Delta X_n)$ obtained from SSA trajectories with randomly sampled initial
conditions.

Using these data, we train conditional generative models for the stochastic updates
\[
\Delta x_{i,n} \approx \Phi_i(x_{1,n},x_{2,n},x_{3,n},x_{4,n},\omega_i), 
\qquad i=1,\dots,4,
\]
and apply the proposed dependence-discovery procedure to identify the state variables on which each update depends.

The variable-removal results and the discovered dependence structure are shown in Figure~\ref{fig:chemical}. The method successfully recovers the correct state dependence implied by the reaction network. Since the reaction propensities depend on $x_1$, $x_2$, and $x_3$, while $x_4$ does not enter any propensity, the reference dependence structure excludes $x_4$ from the fixed-time increments. The recovered structure is consistent with this reference.

We follow the parameter setting, initialization, and time span $T=50$ from \cite{higham2008modeling}, except that we set $c_2=0.1$, and simulate the Michaelis--Menten system independently for $2{,}000$ SSA trajectories. We use an FNN with two hidden layers of $64$ neurons each and ReLU activation function to parameterize the velocity field of the flow matching model, with batch size $5{,}000$ and number of epochs $1{,}000$. In the second stage, we use $M=5{,}000$ distilled samples to fit the GP surrogate. We use the RBF kernel with lengthscale $1.0$ for both the state and latent variables.

\subsection{Hidden fast variables do not obscure local slow-variable dependencies}

We consider the multiscale Lorenz 96 system
\cite{lorenz1996predictability, baptista2024learning}, which couples $m$ slow
variables $x_j$ with $J$ fast variables $y_{k,j}$ for each slow variable:
\begin{subequations}\label{eq:lorenz_96}
\begin{align}
\frac{dx_j}{dt} 
&=  (x_{j+1} - x_{j-2})x_{j-1} - x_j + F
- \frac{hc}{b}\sum_{k=1}^{J} y_{k,j},
\qquad j=1,\ldots,m,
\\
\frac{dy_{k,j}}{dt}
&= -cby_{k+1,j}(y_{k+2,j}-y_{k-1,j}) - c y_{k,j}
+ \frac{hc}{b}x_j,
\qquad k=1,\ldots,J.
\end{align}
\end{subequations}
Here, the fast variables are treated as unobserved nuisance variables. We are therefore interested in the dynamic dependence of each slow variable on the other observed slow variables.

For $m=10$ and $J=10$ in a weak-coupling regime with $h=0.1$, the method recovers
\[
\{x_{j-2},x_{j-1},x_j,x_{j+1}\}
\]
as the dependence set for $\Delta x_j$, consistent with the explicit functional structure of the slow-variable equation. The effect of the unobserved fast variables is encoded in the learned conditional law of $\Delta x_j$ given the observed slow variables, rather than being spuriously attributed to distant slow variables. To assess scalability, we vary $m=10,20,\ldots,90,100,150,200$. The method recovers the correct dependence sets across all values of $m$. With a fixed number of data points $N$, the log-log slope of computational time versus $m$ is $1.81$, consistent with the computational cost of the discovery procedure, $\mathcal{O}(mN^3+m^2N^2)$. In these experiments, the dependence on $x_j$ is supplied as prior knowledge, while the method identifies the remaining variables in $\{x_i:i\neq j\}$ on which $\Delta x_j$ depends.

\begin{figure}[ht!]
    \centering
    \includegraphics[width=0.5\linewidth]{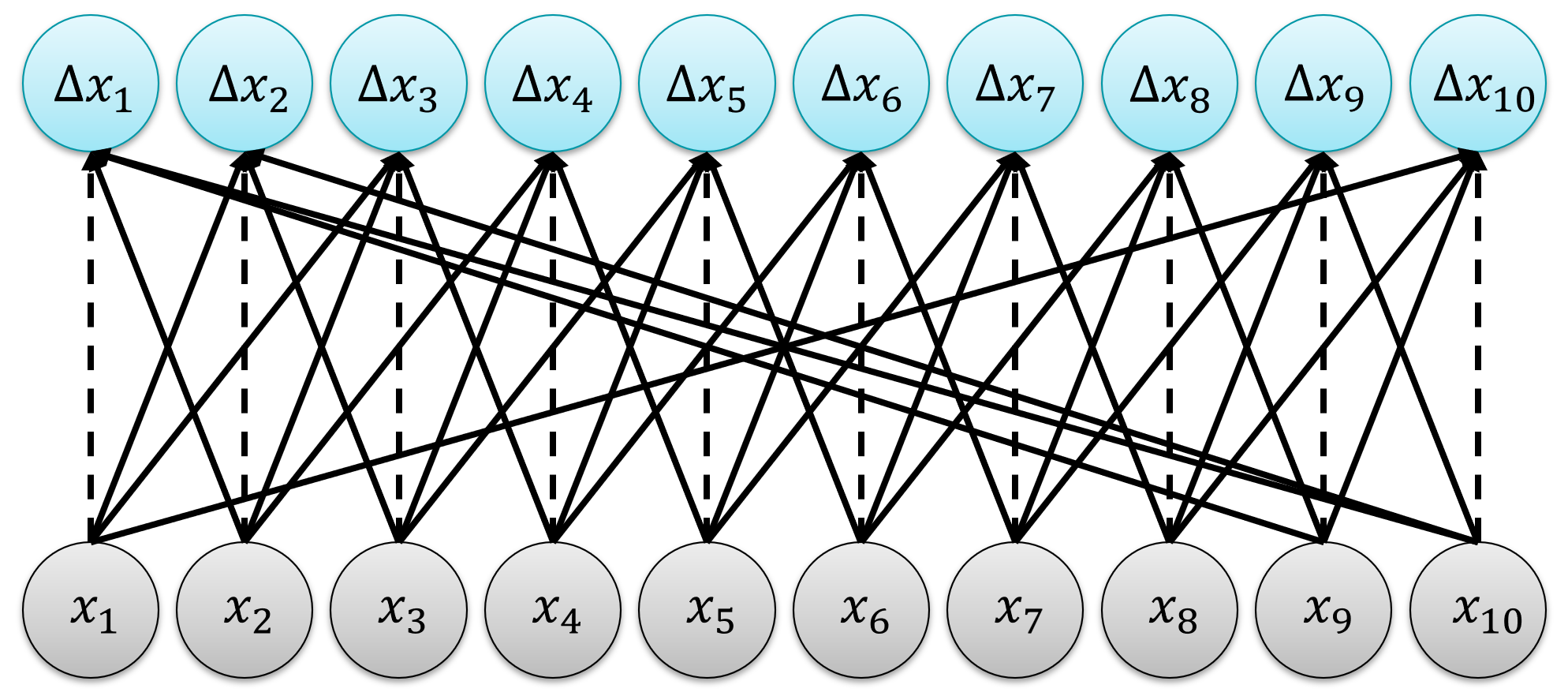}
    \caption{Recovered dependence structure of the multiscale Lorenz 96 system with hidden fast variables. Dashed lines indicate the dependence of $\Delta x_j$ on $x_j$, which is supplied as prior knowledge to the method.}
    \label{fig:multiscale}
\end{figure}

We consider $J=10$ and $m=10,20,\ldots,90,100,150,200$. We use the standard parameters $F=10$, $c=10$, $b=10$, and coupling strength $h=0.1$.
To generate data, we solve \eqref{eq:lorenz_96} using a fourth-order Runge--Kutta scheme with time step $\Delta t=0.001$. The initial conditions are small random perturbations, sampled from Gaussian distributions with mean zero and standard deviation $0.01$, around $F$ for the slow variables and around $0$ for the fast variables. One trajectory is simulated for $102{,}000$ steps; the first $2{,}000$ steps are discarded as burn-in, and snapshots are saved every $10$ steps, yielding $10{,}000$ state-increment pairs.

For $m=10$, we model the conditional distributions of $\Delta x_{j,n}$, $j=1,\ldots,m$, separately by applying the proposed method independently to each target (Figure~\ref{fig:multiscale}). For $m>10$, where the scalability of the method is assessed, we consider only the conditional distribution of $\Delta x_{1,n}$, since the remaining targets have the same computational scaling. In the first stage, we use an FNN with four hidden layers of $64$ neurons each and SiLU activation function to parameterize the velocity field, with batch size $1{,}000$ and number of epochs $2{,}000$. In the second stage, we use $M=2{,}000$ distilled samples and the RBF kernel with lengthscales $5.0$ and $1.0$ for the state and latent variables, respectively.

\subsection{Controlling a stochastic pendulum with missing variables}\label{sec:pendulum}

\begin{figure}[ht!]
    \centering
    \includegraphics[width=1.0\linewidth]{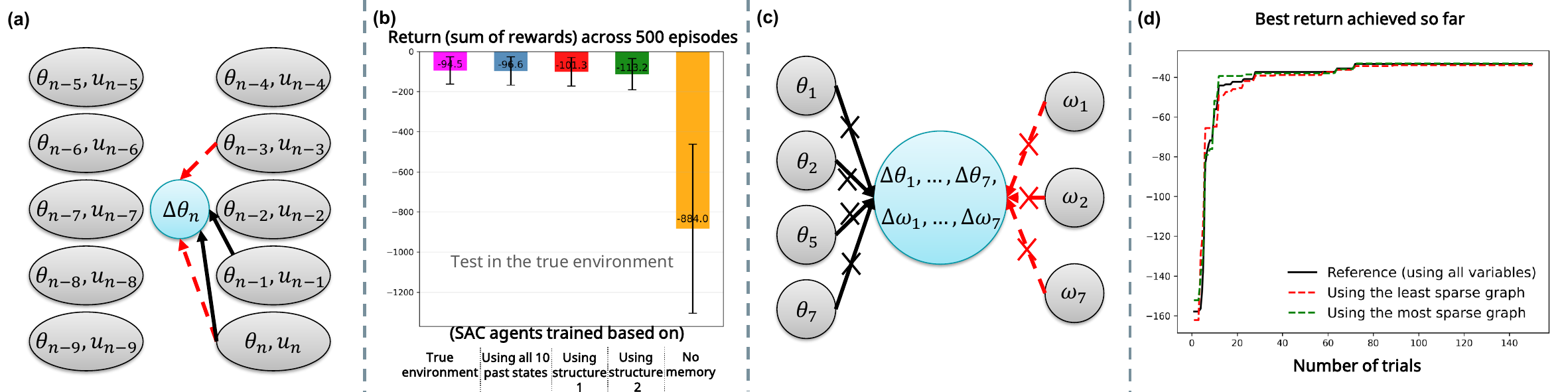}
    \caption{Control examples. (a)--(b): Missing-variable pendulum: memory dependence discovery and policy transfer. (c)--(d): 7-DOF reacher: discovered sparse variables and PETS performance.}
    \label{fig:control}
\end{figure}

We consider a missing-variable control problem in which the transition law of a stochastic system depends on variables encoding memory that are unknown \textit{a priori} and must be discovered from data. The underlying system is a torque-controlled pendulum with multiplicative noise governed by
\begin{equation}\label{eq:pendulum}
    \ddot{\theta}_t
    =
    \frac{3g}{2l}(1+w_t)\sin\theta_t
    +
    \frac{3}{ml^2}u_t,
    \qquad
    w_t\sim\mathcal{N}(0,\sigma^2),
\end{equation}
where $\theta_t$ is the pendulum angle and $u_t$ is the applied torque.

We consider the observation history
\[
(\theta(t),\theta(t-\Delta t),\dots,\theta(t-9\Delta t)),
\]
while the angular velocity $\dot{\theta}_n:=\dot{\theta}(n\Delta t)$ is unobserved. Consequently, the transition law of $\Delta\theta_n$ is not Markovian in $(\theta_n,u_n)$ alone, and past angles and controls $(u_t,\dots,u_{t-9\Delta t})$ are included as candidate lagged inputs. We apply the proposed method to identify a sufficient set of past angles and controls.

Across independent runs, the method identifies two candidate sparse input sets:
\begin{equation}
    \begin{cases}
        \{(\theta_n,u_n),(\theta_{n-1},u_{n-1})\}, & \text{structure 1},\\
        \{(\theta_n,u_n),(\theta_{n-3},u_{n-3})\}, & \text{structure 2}.
    \end{cases}
\end{equation}
See Figure~\ref{fig:control}(a). This reflects structural ambiguity in finite observational data: multiple representations of memory may be sufficient to describe the same transition law. 

Since there is no unique ground-truth sparse input set for this partially observed system, we evaluate sufficiency through the downstream control task of maintaining the upright equilibrium. For each discovered structure, we train a flow matching surrogate and use it as the environment for a SAC agent \cite{haarnoja2018soft}, which is then transferred to the true environment without further training. As shown in Figure~\ref{fig:control}(b), both discovered structures yield agents that transfer successfully, with structure 1 performing marginally better. By contrast, an agent trained using only the input variables $(\theta_n,u_n)$ fails, confirming that memory is necessary. Thus, although the discovered memory representation is not unique, the sparse input sets identified by the method are sufficient for effective downstream control.

To generate data, we collect transition data from \eqref{eq:pendulum} by discretizing the system with time step $\Delta t=0.05$, torque $u_n\in[-2,2]$, mass $m=1$, length $l=1$, gravity $g=9.81$, and noise scale $\sigma=0.3$. We use a customized Gymnasium-compatible simulator \cite{towers2024gymnasium} with the dynamics in \eqref{eq:pendulum}, which integrates semi-implicit Euler updates with angular velocity clipped to the maximum speed $\pm8$. We run $500$ episodes of length $T=200$ from random initial angles $\theta_0\sim\mathrm{Unif}([-\pi,\pi])$ and angular velocities $\dot{\theta}_0\sim\mathrm{Unif}([-1,1])$, applying torques $u_t\sim\mathrm{Unif}([-2,2])$. Each sample records $(\theta_n,u_n,\theta_{n+1})$ with unwrapped $\theta$; for learning, we use inputs $(\cos\theta_n,\sin\theta_n,u_n)$ and target $\Delta\theta_n=\theta_{n+1}-\theta_n$, yielding $100{,}000$ state-increment pairs.

To discover the dependence structure induced by the missing angular velocity from the ten past state-action pairs $\{(\theta_{n-k},u_{n-k})\}_{k=0}^9$, we first use an FNN with four hidden layers of $64$ neurons each and SiLU activation function to parameterize the velocity field, with batch size $1{,}000$ and number of epochs $1{,}000$.
We then use $5{,}000$ distilled samples and the periodic kernel
\[
k_x(x,x^\prime)
=
\exp\left(
-\frac{2\sin^2((x-x^\prime)/2)}{l_x^2}
\right)
\]
with lengthscale $1.0$ for the state variables, an RBF kernel with lengthscale $1.0$ for the control variables, and an RBF kernel with lengthscale $10.0$ for the latent variables. In the second stage, we replace $\cos(\theta_{n-k})$ and $\sin(\theta_{n-k})$ with
\[
\theta_{n-k}
=
\mathrm{atan2}
\left(
\sin(\theta_{n-k}),
\cos(\theta_{n-k})
\right).
\]

As discussed above, we evaluate the sufficiency and accuracy of the discovered structures through downstream control performance. This setup reflects a common scenario in which the true dynamics are unknown and a surrogate must be learned from data. Such a surrogate can then serve as the environment for control methods such as MPC or reinforcement learning (RL), where it is used as a simulator for policy optimization. In this example, we use model-free RL to evaluate the surrogates learned from the discovered sparse input sets.

Specifically, we train a conditional flow matching model for each discovered structure to learn the conditional distribution of $\Delta\theta_n$. We then use the trained flow matching model as a surrogate environment to train a SAC agent~\cite{haarnoja2018soft} using Stable-Baselines3~\cite{stable-baselines3}, with custom environments implemented in Gymnasium~\cite{towers2024gymnasium}. The agent observes a fixed window of $W^\prime=3$ past angles and actions to compensate for the unobserved angular velocity. The trained policy is transferred to the true stochastic environment without further training and evaluated using the mean episode reward over $500$ independently executed episodes. A surrogate based on a sufficient sparse input set should produce a policy that transfers successfully, whereas one based on an insufficient input set should fail.

\subsubsection{Details of the downstream control setup}

The control objective is to swing up and stabilize a stochastic, partially observed pendulum.
The controller applies torque to drive the pendulum toward the upright equilibrium $\theta = 0 \pmod{2\pi}$. The agent does not observe the angular velocity $\dot\theta$; instead, it receives a short history of angle encodings and past actions.
This setting models control under latent velocity, where the policy must infer dynamical information from delayed observations.

The true environment follows a discretized torque-controlled pendulum with multiplicative gravitational noise. At each step $n$, with latent state $(\theta_n,\dot\theta_n)$ and control input $u_n$,
\begin{align}
    w_n &\sim \mathcal{N}(0,\sigma^2), \qquad \sigma = 0.3, \\
    \alpha_n &= \frac{3g}{2l}(1+w_n)\sin(\theta_n) + \frac{3}{ml^2}u_n, \\
    \dot\theta_{n+1} &= \mathrm{clip}\!\left(\dot\theta_n+\alpha_n\Delta t,\,-\dot\theta_{\max},\,\dot\theta_{\max}\right),
    \qquad \dot\theta_{\max}=8~\mathrm{rad/s}, \\
    \theta_{n+1} &= \theta_n+\dot\theta_{n+1}\Delta t.
\end{align}
The noise term $w_n$ perturbs the gravitational acceleration, yielding stochastic transitions even under deterministic policies.

The agent does not receive $\dot\theta_n$. Instead, at step $n$ it observes a sliding window of the most recent $W'=3$ frames. Each frame consists of $(\cos\theta,\sin\theta,u)$, where $u$ denotes the previously applied torque and is set to zero during warm-up. Frames are stacked with the most recent frame first, giving
\begin{equation}
    o_n =
    \big[
        \cos\theta_n,\ \sin\theta_n,\ u_{n-1},\
        \cos\theta_{n-1},\ \sin\theta_{n-1},\ u_{n-2},\
        \cos\theta_{n-2},\ \sin\theta_{n-2},\ u_{n-3}
    \big]
    \in \mathbb{R}^{3W'}.
\end{equation}
Thus, $\dim(o_n)=9$ when $W'=3$. Using $(\cos\theta,\sin\theta)$ avoids discontinuities at $\pm\pi$.
This defines a partially observed Markov decision process: the latent state is $(\theta_n,\dot\theta_n)$, while the policy $\pi(u_n\mid o_n)$ acts only on the 9-dimensional observation.

The control input is a scalar continuous torque applied at each control step after clipping.
Our experiments use the \emph{angle-only} reward, which is fully determined by quantities accessible from the observation history:
\begin{equation}
    r_n = -\left(\bar\theta_{n+1}^2 + 0.001\,u_n^2\right),
\end{equation}
where the wrapped angle is
\begin{equation}
    \bar\theta = ((\theta+\pi)\bmod 2\pi)-\pi \in [-\pi,\pi).
\end{equation}
This reward penalizes deviation from the upright target and control effort. The agent maximizes the finite-horizon return
\begin{equation}
    G=\sum_{n=0}^{T-1}r_n,
\end{equation}
where $T=200$ is the total number of control steps. During evaluation, we report the episode return $G$ over multiple random seeds.

Each episode proceeds as follows: (1) sample the latent initial state $\theta_0\sim\mathrm{Unif}([-\pi,\pi])$ and $\dot\theta_0\sim\mathrm{Unif}([-1,1])$; (2) apply a zero-control warm-up by rolling the true dynamics forward for $K=10$ steps with $u=0$, thereby filling the observation buffer before control begins; and (3) execute the learned policy for up to $T=200$ control steps.
Episodes terminate at the time limit; there is no early termination upon reaching the goal. The warm-up protocol is shared across training and evaluation so that all policies face the same initial observation distribution.

Policies are trained with SAC implemented in Stable-Baselines3~\cite{stable-baselines3} (version~2.7.1), using the default \texttt{MlpPolicy} and the following hyperparameters: learning rate 3e-4, replay buffer size $200{,}000$, batch size $256$, soft-update coefficient $0.005$, discount factor $0.99$, $4$ parallel environments, and $200{,}000$ training timesteps. The checkpoint with the highest evaluation return is retained for testing, with evaluation performed every $10{,}000$ steps over $10$ episodes. The same RL setup is used for true-environment and flow matching surrogate training.

We compare two training regimes: true-environment training and flow matching surrogate training. In the former, SAC interacts directly with the stochastic pendulum dynamics and serves as a reference. In the latter, SAC interacts with a learned simulator that replaces the true transition model.
At each step, a conditional flow matching model samples an angular increment $\Delta\theta$ given a conditioning vector built from past values of $\sin\theta$, $\cos\theta$, and $u$. The surrogate state update is
\begin{equation}
    \theta_{n+1}=\theta_n+\Delta\theta_n,
\end{equation}
where $\Delta\theta_n$ is sampled from the trained flow matching model.
Importantly, the \emph{agent observation and reward remain identical} to those in the true environment ($W'=3$); only the transition model used during training changes.

Different surrogate models correspond to different dependence structures:
\begin{enumerate}
    \item \textbf{Full history:}
    $\sin\theta$, $\cos\theta$, and $u$ at lags $n,n-1,\ldots,n-9$ ($30$ dimensions);
    \item \textbf{Structure 1:}
    $\sin\theta$, $\cos\theta$, and $u$ at lags $n,n-1$ ($6$ dimensions);
    \item \textbf{Structure 2:}
    $\sin\theta$, $\cos\theta$, and $u$ at lags $n,n-3$ ($6$ dimensions);
    \item \textbf{Markov:}
    $\sin\theta$, $\cos\theta$, and $u$ at lag $n$ only ($3$ dimensions).
\end{enumerate}
In all cases, the policy still observes the same 9-dimensional window; the difference lies only in how the surrogate advances the hidden angle state during training. Surrogate episodes also begin with the same true-dynamics zero-control warm-up ($K=10$ steps).

All trained policies, regardless of the simulator used during training, are evaluated on the \emph{same} true stochastic pendulum with $W'=3$, the same reward, $\sigma=0.3$, and the shared warm-up protocol. For each evaluation seed $s$, the environment is reset deterministically from $s$, and the policy acts in deterministic mode using the mean action. Performance is measured by the episode return $G_s$. This \emph{train-on-surrogate/test-on-truth} protocol isolates how well each learned dynamics model supports transfer of control policies to the true system.

\subsection{Sparse probabilistic surrogates preserve model-based control performance}

We apply the proposed method to the 7-degree-of-freedom (7-DOF) reacher problem
\cite{chua2018deep}, a high-dimensional model-based RL benchmark in which a
simulated PR2 robot arm moves its end-effector to a target in 3D space.
Although the MuJoCo simulator dynamics are deterministic, the probabilistic
ensembles with trajectory sampling (PETS) framework \cite{chua2018deep}
models the transition dynamics probabilistically, so that the learned
transition law represents epistemic uncertainty arising from finite data.
This provides a natural setting for studying the input dependence structure
of a learned probabilistic surrogate.


The reacher task involves a simulated PR2 robot arm that must move its end-effector to a randomly placed target in 3D space \cite{chua2018deep}. The true transition dynamics are deterministic and governed by MuJoCo's rigid-body simulation \cite{todorov2012mujoco}. Following PETS, however, the transition dynamics are modeled probabilistically as
\[
P(q_{t+\Delta t}\mid q_t,u_t),
\]
where the stochasticity represents epistemic uncertainty rather than inherent randomness in the simulator. The learned probabilistic dynamics model is used inside MPC to evaluate candidate action sequences before actions are executed in the true environment. 
Chua et al.~\cite{chua2018deep} showed that probabilistic learned dynamics models can outperform deterministic learned dynamics models on this task, motivating the use of a probabilistic surrogate even though the underlying simulator is deterministic.

The observed arm
state is
\[
q(t)=(\theta(t),\omega(t)),
\]
consisting of $7$ joint angles and $7$ angular velocities, and the control
$u(t)\in\mathbb{R}^7$ consists of joint torques. The target variables are the
$14$ state increments
\[
\Delta q_i,\qquad i=1,\ldots,14,
\]
and the candidate inputs are the $21$ state-action variables $(q_t,u_t)$.
The discovery target is therefore the conditional transition surrogate, rather
than the reward function or MPC objective.

\subsubsection{Dependence discovery}

We use the publicly released PETS codebase \cite{chua2018deep} for the 7-DOF
reacher environment and data-generation protocol. Data used for dependence
discovery are generated by rolling out the reacher environment under uniformly
random actions for $100$ independent rollouts. At the start of each rollout,
the environment is reset with a deterministic seed offset by the rollout index;
the target position is sampled by adding independent Gaussian noise with
standard deviation $0.1$ to the target coordinates. Each rollout lasts $150$
environment steps, yielding $15{,}000$ transition samples in total. At each
step, a 7-dimensional action is sampled from the environment action space,
bounded by the MuJoCo motor limits $[-20,20]$, and the simulator advances with
frame skip $2$ and timestep $0.01$.

The raw simulator observation is $17$-dimensional, consisting of the $7$ joint
angles, the $3$ target coordinates, and the $7$ joint angular velocities. For
dependence discovery, we restrict the candidate transition inputs to the $14$
arm-state variables and the $7$ actions; the target location is recorded and
used by the task cost, but is not treated as a candidate input to the transition
surrogate. For every transition, the dataset stores the current raw observation,
action, next raw observation, reward, goal position, and observation change.

We model the joint conditional distribution of the \(14\)-dimensional state increment given the \(21\) state-action inputs using a flow matching model. The velocity field is parameterized by an FNN with four hidden layers of \(128\) neurons and SiLU activation function, trained with batch size \(1{,}000\) for \(1{,}000\) epochs. In the second stage, we use \(2{,}000\) distilled samples and an RBF kernel with lengthscale \(10.0\) for both the state-action and latent variables. 

Using the resulting joint generative model, we perform dependence analysis separately for each of the \(14\) output coordinates. To construct a single sparse input set for the PETS surrogate, we take the union of the variables retained across all \(14\) coordinates, so that a variable is kept if it contributes to the conditional law of at least one transition coordinate. We repeat the experiment five times independently to assess run-to-run variability.

Across the five runs, the method consistently removes
$\theta_1,\theta_2,\theta_5,\theta_7$. Run-to-run variability occurs for
$\omega_1,\omega_2,\omega_7$, which are removed in some runs but retained in
others. As shown in Figure~\ref{fig:control}(c), the least sparse structure
removes only
\[
\{\theta_1,\theta_2,\theta_5,\theta_7\},
\]
while the most sparse structure additionally removes
\[
\{\omega_1,\omega_2,\omega_7\}.
\]

\subsubsection{Downstream model-based control}

We evaluate the discovered structures within the original PETS model-based RL
pipeline. All reacher experiments use the publicly released PETS implementation
and experimental protocol of \cite{chua2018deep}; we refer readers to that work
for details of the probabilistic dynamics model, trajectory-sampling scheme,
MPC planner, and cross-entropy method (CEM) optimization.

In all comparisons, we use the standard PETS probabilistic dynamics model and
change only the input coordinates supplied to it. The full baseline receives
all $21$ state-action inputs, whereas the sparse models receive only the
variables retained by the discovered structures. The output space, planning
objective, cost function, action bounds, data-collection schedule, and MPC/CEM
procedure are kept fixed across all models. Thus, the comparison isolates the
effect of changing the input structure of the probabilistic transition
surrogate.

Both the least and most sparse surrogates achieve performance comparable to the
full PETS baseline, as shown in Figure~\ref{fig:control}(d), where return denotes
the cumulative task reward and the plotted curve reports the mean best return
achieved so far across trials. For visualization, we increase the number of
trials from $100$ to $150$ and plot the average over three independent
experiments. These results show that substantial reduction of the surrogate
input space can preserve the reported model-based control performance.

\subsubsection{Discussion of the result}

The discovered structures are data-dependent: the results reported here are
based on $100$ random-control rollouts used for dependence discovery, and a
different dataset may yield different estimated structures. PETS also operates
as an iterative loop: starting from random-control data, a probabilistic
dynamics model is fitted and used for MPC planning; the resulting trajectories
are then added to the dataset, and the model is refined. As this loop
progresses, the data distribution shifts from random exploration toward
task-relevant, model-guided trajectories.

Consequently, the dependence structure of the learned probabilistic surrogate
may also change across PETS iterations. Variables with weak dependence under
random controls may become important in task-critical regions of the
state-action space, while variables retained under broad random exploration may
become unnecessary once the data concentrate near successful trajectories. A
natural extension of the proposed approach is therefore to repeat dependence
discovery periodically as the dataset is updated, allowing the sparse input
structure to evolve alongside the dynamics model. This would yield an adaptive
pipeline in which dependence discovery and model refinement are interleaved.

\end{document}